\documentclass[a4paper,11pt]{article}
\pdfoutput=1
\usepackage{jheppub}
\usepackage{graphicx}
\input{epsf}
\usepackage{epsfig}
\usepackage{epstopdf}
\usepackage{calligra}
\usepackage{subcaption}
\usepackage{color}
\usepackage[normalem]{ulem}

\usepackage{color} 
\usepackage[section]{placeins}
\usepackage{float}
\usepackage{graphicx,wrapfig,hyperref}
\usepackage{slashed}
\usepackage{tikz-cd} 

\usepackage{caption}
\DeclareCaptionFont{10pt}{\fontsize{10pt}{12pt}\selectfont}
\newcommand{\ra}{\rightarrow} 
\newcommand{\Ra}{\Rightarrow}

\newcommand{\mn}{{\mu\nu}}

\newcommand{\equ}[1]{\begin{equation} #1 \end{equation}}
\newcommand{\ali}[1]{\begin{align} #1 \end{align}}
\newcommand{\p}{\partial}

\newcommand{\vev}[1]{\langle #1 \rangle}
 
\DeclareMathOperator\csch{csch} 
\DeclareMathOperator\tr{tr}

\def\ellK{\ell} %\ell_K
\def\ellads{\ell} 
\def\tads{{\rm{t}}} 
\def\zads{{\rm{z}}}
\def\Sjac{{\text{\large \rm{S}}}}

\title{The gravitational dynamics of kinematic space}
\author[1,2]{Nele Callebaut}

\affiliation[1]{Department of Physics and Astronomy, Ghent University, Krijgslaan 281-S9, 9000 Gent, Belgium}

\affiliation[2]{Department of Physics, Princeton University, Princeton, NJ 08544, USA}

\emailAdd{nelec@princeton.edu} 

\arxivnumber{1808.10431}

\abstract{
We show that the dynamics of the kinematic space of a 2-dimensional CFT is gravitational and described by Jackiw-Teitelboim theory. 
We discuss the first law of this 2-dimensional dilaton gravity theory to support the relation between modular Hamiltonian and dilaton that underlies the kinematic space construction. It is further argued that Jackiw-Teitelboim gravity can be derived from a 2-dimensional version of Jacobson's maximal vacuum entanglement hypothesis. Applied to the kinematic space context, this leads us to the statement that the kinematic space of a 2-dimensional boundary CFT can be obtained from coupling the boundary CFT to JT gravity through a maximal vacuum entanglement principle.    
}

\begin{document}
\maketitle

\section{Introduction and overview}

Kinematic space has been defined as the space of intervals on a constant time slice of a given 2-dimensional CFT \cite{Czech:2015qta,deBoer:2015kda}, or the space of pairs of points \cite{Czech:2016xec,deBoer:2016pqk}. It has the structure of the product of two 2-dimensional de Sitter spaces, corresponding to the left-moving and right-moving sector of the CFT. We will restrict to cases where it  
is simply equal to the diagonal de Sitter. When the CFT is holographic, kinematic space can also be referred to as the space of corresponding boundary-anchored 
geodesics of the AdS bulk and has been used as a tool to study the induced dynamics of the AdS bulk \cite{Czech:2016tqr}. In contrast, we are interested here in the dynamics of the kinematic space itself, which we will discuss to be the `dynamics' of 2-dimensional gravity. 
More precisely, we will give an interpretation of kinematic space as a theory of Jackiw-Teitelboim (JT) gravity.  
The results in this paper are complementary to the discussion of the entropic origin of JT gravity in \cite{Callebaut:2018nlq}, focusing on 
two new aspects: kinematic space and the maximal entanglement principle. 

In section \ref{LiouvilleKsection} we summarize the original kinematic space construction of \cite{Czech:2015qta}, with an emphasis on the role of the one-interval entanglement entropy of the CFT as a Liouville field, which was first pointed out in \cite{deBoer:2016pqk}.  
The original definition for the metric on kinematic space in terms of entanglement \cite{Czech:2015qta} is then just the Liouville metric, given in equation \eqref{ds2K}. The entanglement of the (2-dimensional) CFT becomes a metric field in the kinematic space construction, and the `entanglement dynamics' of the CFT, or in other words the dynamics of kinematic space,  should then naturally be described  by a theory of (2-dimensional) gravity. While pure Einstein gravity is trivial in two dimensions, the Jackiw-Teitelboim theory we will encounter in section \ref{JTKsection}    
is a theory of dilaton gravity, with not just a metric but also a dilaton field.  
We end section \ref{LiouvilleKsection} with  
the observation that the Liouville stress tensor for the entanglement is given by the vacuum expectation value of the CFT stress tensor evaluated at the interval endpoints and comments on the bulk AdS$_3$ perspective.

It was  
observed in \cite{deBoer:2015kda} that the one-interval entanglement perturbations $\delta S$ obey a de Sitter Klein-Gordon equation on the kinematic space ($K$). This constitutes one of four Jackiw-Teitelboim equations of motion for a dilaton $\delta S$, 
in a conformal gauge determined by the entanglement $S$. The Liouville equation for $S$ is another, and we complete the picture of $\delta S$ obeying JT dynamics on kinematic space by showing the two remaining constraint equations of motion are satisfied as well. This is done in section \ref{JTKsection}. The `$K$ on-shell identities' in \eqref{EOM1K}-\eqref{EOM4K} are concluded to be imposed as equations of motion by a JT theory that governs the dynamics of $K$. This is the main conclusion of the paper. We can take the identification of the entanglement perturbations with the dilaton in \eqref{Kconstruct} 
as constructing principle of $K$. 

We discuss similar `$K_\p$ identities'  \eqref{EOM1Kb}-\eqref{EOM4Kb} for the boundary kinematic space $K_\p$ of a boundary CFT$_{2}$ \cite{Karch:2017fuh} in section \ref{Kbsection}. While both $K$ and $K_\p$ have a Jackiw-Teitelboim description, there are two main differences with the previous discussion. Firstly, $K_\p$ has an AdS$_2$ geometry, while $K$ has a dS$_2$ geometry. Second, it are the lightcone coordinates of the boundary CFT$_2$ (rather than the interval endpoints in the case of a CFT without boundary) that determine the lightcone coordinates of the associated kinematic space. This difference is readily seen from comparing the metrics \eqref{ds2K} and \eqref{ds2Kb}. 
It is this difference that allows to interpret the construction of the boundary kinematic space of a boundary CFT as the process of coupling that boundary CFT to AdS$_2$ JT gravity. Such an interpretation of the de Sitter kinematic space remains less clear.

Jacobson's maximal entanglement hypothesis \cite{Jacobson:2015hqa} shares a similar set-up and ingredients with the kinematic space discussion. Namely, it considers a CFT on a %maximally symmetric 
background geometry (without gravity) and imposes conditions on the entanglement in the CFT that 
can be interpreted as a prescription to couple the CFT to semi-classical gravity. Based on the 
interpretation of boundary kinematic space in the conclusion of section \ref{Kbsection}, we set out in section \ref{Jacsection} to examine the relation between boundary kinematic space and the maximal entanglement hypothesis applied to a 2-dimensional boundary CFT. To this end we review the original argument of \cite{Jacobson:2015hqa}, valid for dimensions greater than two, in section \ref{Jacsectionreview}. Next, we discuss in section \ref{JacsectionJT} the coupling of a 
CFT to JT gravity as a 
2-dimensional application of the Jacobson argument. This involves reformulating 
the JT first law as 
a condition on entanglement in the CFT. 
In section \ref{JacsectionK} we interpret the constructing principle of boundary kinematic space in equation \eqref{Kbconstruct} as such an entanglement condition,  
and claim that the kinematic space of a 2-dimensional boundary CFT can thus be obtained from coupling the boundary CFT to JT gravity through a maximal vacuum entanglement principle.

One confusing aspect of the discussion is the distinction between the CFT on a flat background geometry in the original set-up and the CFT on an (A)dS background geometry in the obtained kinematic space. In section \ref{JTsection} we therefore consider a CFT on an AdS background in JT gravity theory, regarding it as instructive to study this set-up without reference to kinematic space. It can be read as a stand-alone section that discusses entanglement in the JT theory (section \ref{JTsectionS} and \ref{modhamsection}), the JT mass formula and first law (section \ref{JTsectionMass}), and the interpretation of the vacuum contribution to the dilaton as a differential entropy (section \ref{Phi0section}).  
However, each of these discussions pertains to the kinematic space context, in ways discussed at the end of each subsection. 
In particular, the discussion of entanglement and the JT mass formula reveals a natural link between entanglement and metric on one hand, and between modular Hamiltonian and dilaton on the other. 
Because the (local) entanglement and the modular Hamiltonian are given by the same formulas in the boundary CFT coupled to JT and in the flat space boundary CFT,  
this gives us 
more insight into the arbitrary-looking observations  \eqref{EOM1Kb}-\eqref{EOM4Kb} (as well as \eqref{EOM1K}-\eqref{EOM4K}, by analogy). The discussion of the vacuum contribution to the dilaton reveals a relation to the Schwarzian theory and cMERA, which is quite natural in light of the relation between JT theory and kinematic space.  

We conclude with a discussion of the obtained results and possible future directions in section \ref{discussion}.

\section{Kinematic space} 
\label{LiouvilleKsection}

\begin{figure}[t]
	\begin{center}
		\begin{tikzpicture}
		\draw [fill=yellow, opacity=0.2]
		(-2.4,0) -- (0,2.4) -- (2.4,0) -- cycle;	
		\draw[->][thick] (-3,0) -- (3,0); 
		\draw (3.2,0.2) node {\footnotesize $x$};
		\path[->][thick](0,-.5) edge (0,2.5); 
		\draw (0.2,2.7) node {\footnotesize $t$}; 	
		\draw[|-|][color=blue]  (-0.5,0) -- (1.5,0);
		\draw[-][ultra thick,color=blue]  (-0.5,0) -- (1.5,0); 
		\draw (-0.5,-0.25) node {\footnotesize $u$};
		\draw (1.5,-0.25) node {\footnotesize $v$};  
		\draw[dashed] (-0.5,0) -- (0.5,1);
		\draw[dashed] (0.5,1) -- (1.5,0);
		\draw[fill=black] (0.5,1) circle (0.08);
		\end{tikzpicture}
	\end{center}
	\vspace{-2mm}
	\caption{Interval $x \in [u,v]$ (in blue) on a constant time slice of a CFT$_2$ in vacuum state $|0\rangle_X$. The kinematic space construction involves 1) promoting the $t=0$ time-slice of the CFT to past infinity of kinematic space, 
		2) identifying the interval endpoints $u$ and $v$ with kinematic space lightcone coordinates, and 3) using the one-interval entanglement formula to define a hyperbolic metric on kinematic space through \eqref{ds2K}. The yellow triangle is 
		a sketch of the emergent dS$_2$ kinematic space, superimposed here on the picture of the CFT background.}  \label{figK}
	\vspace{-3mm}
\end{figure}
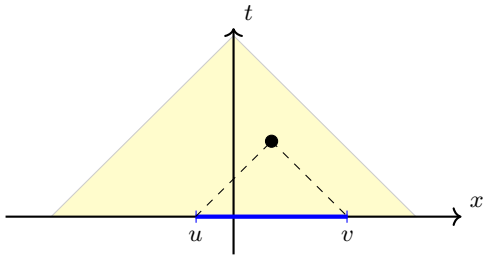

Consider a CFT on a 2-dimensional Minkowski geometry $ds^2 = -dx^+ dx^-$ with lightcone coordinates $x^\pm = t \pm x$ and large central charge $c$. We take the theory to be in the vacuum state 
\ali{
	|0\rangle_X. 
}
This notation refers to the state that contains no quanta that are positive frequency with respect to time $\frac{X^++X^-}{2}$, for lightcone coordinates $X^\pm$ that are related to $x^\pm$ by a general conformal transformation 
\ali{
	(x^+,x^-) \ra (X^+(x^+), X^-(x^-)).  
}
The state is characterized by a stress tensor expectation value\footnote{
	Here the curly brackets denote the Schwarzian derivative, defined as $\{ f, x \} = \frac{f'''}{f'} - \frac{3}{2} \left(\frac{f''}{f'}\right)^2$.
} 
\ali{
	\vev{T_{x^\pm x^\pm}} = -\frac{c}{24 \pi}  \{ X^\pm, x^\pm \}  \qquad \qquad
	\label{vacstatedef}
}       
that vanishes in the frame of an $X^\pm$-observer, who measures a stress tensor $T_{X^+X^+}$ related to $T_{x^+x^+}$ by  
\ali{
	T_{x^+x^+} = -\frac{c}{24\pi} \{ X^+,x^+ \} + T_{X^+X^+} \left(\frac{dX^+}{dx^+}\right)^2.  
}
The $X^\pm$ coordinates are the `uniformizing coordinates' of the CFT. 
We will moreover restrict to states with
\ali{
	X^+=X^-=X, 
} 
which have equal right-moving and left-moving stress tensor components 
\ali{
	\vev{T_{x^-x^-}} = \vev{T_{x^+x^+}} = -\frac{c}{24 \pi}  \{ X, x^\pm \}.  
} 
We can then use the vacuum formula for the entanglement of an interval in the CFT to write 
\ali{
	S(U,V) &= \frac{c}{12} \log \frac{(V - U)^2}{\delta_U^2}, \qquad V = X^+(v), \,\, U = X^+(u)  
}
for the contribution of right-moving degrees of freedom to the entanglement, which functionally depends on the (transformed) interval endpoints $U$ and $V$, with $\delta_U$ a UV cutoff in $X$ coordinates. Because the cutoff transforms non-trivially under the conformal transformation $X(x)$, the entanglement as a function of the interval endpoints $u$ and $v$ (see figure \ref{figK}) is given by \cite{Holzhey:1994we}  
\ali{	
	S(u,v) &= \frac{c}{12} \log \frac{(X(v) - X(u))^2}{\delta_u^2 \, {X}'(v) {X}'(u)}.  \label{ent}	
}
It immediately follows from this expression that $S$ satisfies 
\ali{
	\p_u \p_v \left( \frac{12}{c} S \right) = \frac{2}{\delta_u^2} e^{-\frac{12}{c} S}.   \label{LiouvilleKeq}
}
Under the identification 
\ali{
	\frac{12}{c} S = -\omega_u + 2 \log \frac{2 \ellK}{\delta_u},   \label{Sofomega}
}
the equation \eqref{LiouvilleKeq} can be recognized as the classical Liouville equation 
\ali{ 
	4 \p_u \p_v \omega_u + \Lambda e^{\omega_u} = 0 
} 
for a Liouville field $\omega$ (we will not always explicitly write the subindex referring to the coordinate system), expressing constant curvature  	
\ali{ 
	R = \Lambda, \qquad \Lambda = \frac{2}{\ell^2} 
} 	
of the Liouville metric 
\ali{
	ds^2_K &= e^{\omega_u} du dv = \left(\frac{2\ellK}{\delta_u}\right)^2 e^{-\frac{12}{c} S(u,v)} du dv \label{ds2K}.  
}
On the solution \eqref{ent}, the Liouville metric becomes (a slicing of) the 2-dimensional de Sitter metric 
\ali{ 
	ds^2_K &= \frac{4 \ellK^2 dU dV}{(V-U)^2} = \frac{4 \ellK^2 {X}'(u) {X}'(v) du \, dv}{(X(v)-X(u))^2}.	
}
Classical Liouville theory solves the `uniformization problem': given a 2-dimensional manifold with local lightcone coordinates $u,v$, its most general 
metric can be parametrized by the Liouville field $\omega_u$ according to \eqref{ds2K}, and the solution to the Liouville equation \eqref{LiouvilleKeq} lays a hyperbolic metric (in this case dS$_2$) on the manifold. This can always be done, by transforming the lightcone coordinates to `uniformizing coordinates' $U$ and $V$. 
We thus see that the one-interval entanglement of the given 
CFT$_2$ solves the uniformization problem for a 2-dimensional manifold with lightcone coordinates given by the endpoints $u$ and $v$ of the interval. Each point in this manifold labels a CFT interval; it is the space of CFT intervals, which was named `kinematic space'\footnote{
	Note that here we follow the `original' definition of kinematic space as the space of CFT intervals in \cite{Czech:2015qta,deBoer:2015kda}, rather than the more general definition as the space of pairs of points in \cite{Czech:2016xec,deBoer:2016pqk}. The latter 
	makes use of the OPE block structure of the CFT, while the first is  based more directly on the entanglement of the CFT. In this paper we are interested in the kinematic space of  \cite{Czech:2015qta,deBoer:2015kda}, which describes the `entanglement dynamics' of the CFT, as we are interested in to which extent the entanglement of the CFT can be treated as a dynamic field itself. 
	} in \cite{Czech:2015qta}. The kinematic space metric given in \eqref{ds2K} corresponds to the definition $ds^2_K = \frac{4}{\Lambda} \p_u \p_u \left(\frac{12}{c} S \right) \, du \, dv$ of \cite{Czech:2015qta}.

Because of our restriction to states $|0\rangle_X$ with $X^+ = X^-=X$, we focus on the case where the kinematic space of right-moving degrees of freedom equals the one of left-moving degrees of freedom and the general kinematic space with metric dS$_2$ $\times$ dS$_2$ reduces to one, diagonal dS$_2$ \cite{deBoer:2016pqk}. %,Czech:2016xec
The dS$_2$ metric has a boundary at $U=V$ (or $u=v$),  which can be identified with the constant time slice of the CFT to allow a natural association of a \emph{point} in kinematic space $K$ with an interval $[U,V]$ on that time-slice of the CFT. The interval endpoints become lightcone coordinates in $K$. The construction of $K$ is summarized in figure \ref{figK}.

The Liouville stress tensor associated with the Liouville field $\omega$ is \cite{Seiberg:1990eb} 
\ali{
	T_{uu}^L &= -\frac{c}{24\pi} \left(\frac{1}{2} (\p_u \omega)^2 - \p_u^2 \omega \right).  
}
Substituting the relation between $\omega$ and $S$ we find that the Liouville stress tensor for the vacuum entanglement as a Liouville field is given by the CFT stress tensor evaluated at the interval endpoints (see also \cite{Wall:2011kb}) 
\equ{ 
\begin{aligned}  	
	T_{uu}^L &= \frac{1}{2\pi} \left( - \frac{6}{c} (\p_u S)^2 - \p_u^2 S \right) = \frac{c}{24\pi} \{ X^+,u \} = -\vev{T_{x^+x^+}(x^+=u)}     \\
	T_{vv}^L &= \frac{1}{2\pi} \left( - \frac{6}{c} (\p_v S)^2 - \p_v^2 S \right) = \frac{c}{24\pi} \{ X^+,v \} = -\vev{T_{x^+x^+}(x^+=v)}   
\end{aligned}	\quad .  \label{TuuL}
	}

Let us comment on the AdS$_3$ perspective to make contact to another occurrence of Liouville theory in this set-up. If the 
CFT under consideration is holographic, it has an AdS$_3$ dual. 
In 3 dimensions, Einstein-Hilbert gravity with a negative cosmological constant is trivial in the sense that there are no propagating degrees of freedom. All solutions have constant negative curvature and are thus locally AdS$_3$. The most general such solution that is asymptotically AdS$_3$ (asAdS$_3$) is the Banados metric, with radius $l$. In Fefferman-Graham notation: 
\ali{
	ds^2_{Banados} &= l^2 \frac{d\rho^2}{4\rho^2} + l^2 \left( L_-(x^-) (dx^-)^2 + L_+(x^+) (dx^+)^2 \right) - \left(\frac{l^2}{\rho} + l^2 \rho L_+ L_- \right)dx^- dx^+   \nonumber  
}
with the boundary at AdS radius $\rho \ra 0$. The $L$ functions correspond to the Brown-York stress tensor components of asAdS$_3$ gravity
\cite{deHaro:2000vlm}, 
and thus through AdS/CFT with the corresponding expectation values of the CFT stress tensor \cite{Roberts:2012aq}: 
\ali{
	L_\pm 
	= \frac{8\pi G_3}{l} \langle T_{x^\pm x^\pm} \rangle    \label{LisT}
}	
where $G_3$ is the 3-dimensional gravitational constant. 
Because there are no local bulk degrees of freedom, the physics of AdS$_3$ gravity is located at the boundary and different boundary conditions generate different boundary dynamics. 
In particular, the Brown-Henneaux boundary conditions, imposing an asAdS$_3$ metric, yield an asymptotic symmetry algebra given by the Virasoro algebra, the conformal algebra in 2 dimensions, with $c = 3 l/2G_3$.  
The boundary dynamics of asAdS$_3$ gravity are then described at the classical level by a Liouville theory whose stress tensor is such that \cite{Coussaert:1995zp,Bautier:2000mz,Carlip:2005tz}
\ali{
	T^{Liou}_{\pm\pm} 
	= \langle T_{x^\pm x^\pm} (x^\pm) \rangle = \frac{c}{12\pi} L_\pm(x^\pm), \qquad c = \frac{3 l}{2 G_3} .  \label{TLiou}
}
There are thus different Liouville theories at play in the context of AdS$_3$/CFT$_2$. The equations \eqref{TuuL} and \eqref{TLiou} suggest a deeper relation between the Liouville theory associated with kinematic space and the one associated with the AdS$_3$ boundary dynamics, a better understanding of which is left for future work.

\section{JT theory for kinematic space} \label{JTKsection} 

The vacuum modular Hamiltonian for the interval in figure \ref{figK} is defined by writing the reduced density matrix of the system as $\rho = e^{-H_{mod}}/\tr e^{-H_{mod}}$ and    
is given by \cite{Casini:2011kv} 
\ali{
	H_{mod}(U,V) &= 2\pi \int_U^V dS \frac{(S-U)(V-S)}{V-U} T_{X^+ X^+}(S) 
}
or 	
\ali{ 	
	H_{mod}(u,v) &= 2\pi \int_u^v ds \frac{\left( X(s)-X(u) \right) \left(X(v)-X(s)\right)}{\left(X(v)-X(u)\right) X'(s)} \delta T_{x^+ x^+}(s).  \label{HmodCFT}
}
It is normalized such that its vacuum expectation value vanishes, i.e.~the stress tensors in the integral are the (covariantly transforming) vacuum-subtracted ones
\ali{ 
	\delta T_{x^+ x^+}  = T_{X^+ X^+}  \left(\frac{dX^+}{dx^+}\right)^2.
}     
We again focus only on the contribution of right-moving degrees of freedom, equal to the contribution of left-moving ones. 

Upon perturbing the vacuum state slightly to the state $|\psi \rangle$, with reduced density matrix $\rho' = \rho + \delta \rho$, the entanglement entropy $S = -\tr ( \rho \log \rho)$ changes by an amount $\delta S = -\tr (\delta \rho \log \rho)$ to first order in $\delta \rho$ if we make use of $\tr (\delta \rho) = 0$. We can write 
\ali{
	\delta S = \delta \vev{H_{mod}}   \label{FLE}
}
with the notation 
\ali{
	\delta \vev{H_{mod}} = \vev{H_{mod}}_\psi.  
}
This relation is known as the `first law of entanglement'.

The authors of \cite{deBoer:2015kda} were the first to notice that the entanglement perturbations $\delta S$ satisfy a Klein-Gordon equation on an emergent 2-dimensional de Sitter geometry or kinematic space. This provided an alternative definition of kinematic space, that was checked to be equivalent to the definition of \cite{Czech:2015qta} in \cite{Asplund:2016koz}. It was checked for different states of the type \eqref{vacstatedef}, with an emphasis on the thermal one, which has $X = \frac{\beta}{2\pi} \tanh(\frac{2\pi}{\beta} \, x)$. 
The deeper reason for the equivalence is the fact that the entanglement is a Liouville field. Indeed, \eqref{LiouvilleKeq} expresses the constant curvature equation for the metric \eqref{ds2K}, which served as kinematic space definition in \cite{Czech:2015qta}. On the other hand, when linearized ($S \ra S + \delta S$), \eqref{LiouvilleKeq} gives rise to the wave equation on de Sitter,  
$(\Box + \Lambda) \delta S = 0$. The same equation is true for the modular Hamiltonian by the first law of entanglement. 

We complete the set of equations that are satisfied by $S$ and $H_{mod}$ to what we could call the `kinematic space on-shell identities': 
\ali{
	\p_u \p_v \left( \frac{12}{c} S \right) &= \frac{2}{\delta_u^2} e^{-\frac{12}{c} S}  \label{EOM1K} \\ 
	e^{-\frac{12}{c}S} \p_u \left( e^{\frac{12}{c}S} \p_u H_{mod} \right) &= 2 \pi \, \delta T_{x^+ x^+}(x^+=u)  
	\label{EOM2K} \\  
	e^{-\frac{12}{c}S} \p_v \left( e^{\frac{12}{c}S} \p_v H_{mod} \right) &= 2 \pi \, \delta T_{x^- x^-}(x^-=v) 
	\label{EOM3K} \\ 
	\p_u \p_v H_{mod} &= -\frac{2}{\delta_u^2} e^{-\frac{12}{c}S} H_{mod} . \label{EOM4K}
}

By expressing these equations in terms of the natural metric \eqref{ds2K} on $K$, they take the more transparent form 
\ali{
	R &=  \Lambda \label{} \\ 
	\nabla_u \p_u  H_{mod} &= 2\pi \, \delta T_{x^+x^+}(x^+=u)   \label{}\\
	\nabla_v \p_v  H_{mod} &= 2\pi \, \delta T_{x^+x^+}(x^+=u)  \label{}\\
	(\Box + \Lambda) H_{mod} &= 0 \label{} 
}
with $	\Lambda = \frac{2}{\ellK^2}$,  $\Box = \frac{\delta^2}{\ellK^2} e^{-\omega} \p_u \p_v$ and $R = -\frac{\delta^2}{\ellK^2} e^{-\omega} \p_u \p_v \omega$. This set of equations is to be compared to the equations of motion  \eqref{EOM1dS}-\eqref{EOM4dS} of Jackiw-Teitelboim theory, reviewed in appendix \ref{JTreview}.

The first equation is the Liouville equation \eqref{LiouvilleKeq}, the last equation is the dS wave equation and the second and third are Jackiw-Teitelboim constraint equations. Together, these identities map to the Jackiw-Teitelboim equations of motion in conformal gauge \eqref{EOM1dS}-\eqref{EOM4dS}, with the entanglement identified as (minus) the 
Liouville field $\omega$ and the 
modular Hamiltonian as (minus) the  
dilaton (as quantum field operator):
\ali{
	\omega &= - \frac{12}{c}S + 2 \log \frac{2 \ellK}{\delta}, \label{identif0} \\ 
	\Phi &=  - 4G \, H_{mod} + \Phi_0. \label{identif}
}
Here we included a zero-mode term $\Phi_0$ in the solution of the dilaton. Its interpretation will be discussed in section  \ref{Phi0section}.

We can now write a kinematic space JT action (see \eqref{JTaction}) 
\ali{
	I_{JT}[g,\Phi,\phi_m]  = \frac{1}{16 \pi G} \int d^2 \sigma 
	\sqrt{-g} \,  \Phi (R - \Lambda) \, \, + \, \, I_{m}[g,\phi_m]  \label{IJTKaction}
} 
for a metric $ds^2 = g_\mn d\sigma^\mu d\sigma^\nu$ with lightcone coordinates $u$ and $v$, a dilaton $\Phi$ and conformal matter fields $\phi_m$.  
On a solution \eqref{identif0}-\eqref{identif}, the JT equations give rise to the kinematic space on-shell identities \eqref{EOM1K}-\eqref{EOM4K} when the (vacuum-subtracted) CFT stress tensor evaluated at the interval endpoints equals the dilaton matter stress tensor living on kinematic space 
\equ{ 
\begin{aligned} 
	\delta T_{x^+x^+}(x^+=u) = T_{uu}^m(u) \\
	\delta T_{x^+x^+}(x^+=v) = T_{vv}^m(v)  
\end{aligned} \,\, . \label{TmJTdS}
}
That is, the JT theory of $K$ is coupled to a matter CFT on the dS$_2$ kinematic space background so that the above is true. 
We conclude that the entanglement dynamics or kinematic space dynamics of a given 2-dimensional CFT are governed by the JT theory \eqref{IJTKaction} with metric, dilaton and matter fields specified in equations \eqref{identif0}, \eqref{identif} and \eqref{TmJTdS}. 

We can write equation \eqref{identif}, with notation $\Phi_T$ for $\Phi - \Phi_0$, as  
\ali{
	\frac{\Phi_T}{4G} = - H_{mod}. \label{Kprinciple} 
}
We take this identification as a defining principle for the construction of kinematic space: 
associate a point in $K$ with a CFT interval by promoting $H_{mod}$ to a field operator $\frac{\Phi_T}{4G}$ living on kinematic space. Here $G$ is just a dimensionless number, but we include it to make the comparison to JT more direct. 

We have not written explicitly but assume the presence of the Polyakov action \eqref{IPol} in the action \eqref{IJTKaction} to reproduce the $K$ identities after taking the expectation value in the state $|\psi \rangle$. The effect of the trace anomaly can be absorbed in $\Phi_0$ so that the semi-classical dilaton contribution $\vev{\Phi_T}$ maps to a classical $\Phi_T$ (some more details are given in appendix \ref{JTreview}). The constructing principle of the semi-classical JT theory of $K$ is then    $\frac{\Phi_T}{4G} = - \vev{H_{mod}}_{\psi} $ 
or 
\ali{ 
	\frac{\delta \Phi}{4G} = - \delta \vev{H_{mod}}.   \label{Kconstruct}
}
This notation helps us  
keep in mind that the state $|\psi\rangle$ is a small reduced density matrix perturbation away from the vacuum.

\section{JT theory for boundary kinematic space} \label{Kbsection}

\begin{figure}[t]
	\begin{center}
		\includegraphics[]{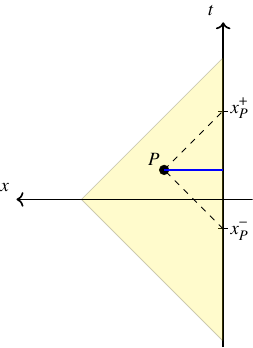} \qquad 
	\end{center}
	\vspace{-2mm}
	\caption{
		Interval [$P$, boundary $x=0$] (in blue) on a constant time slice of a bCFT$_2$ in vacuum state $|0\rangle_X$. It 
		is specified by the location of the point $P$ in CFT lightcone coordinates $(x^+_P,x^-_P)$.   
		The kinematic space construction involves 1) promoting the $x=0$ boundary of the CFT to spacelike infinity 
		of kinematic space, 2) identifying 
		$x^+$ and $x^-$ with kinematic space lightcone coordinates, and 3) using the one-interval entanglement formula to define a hyperbolic metric on kinematic space through equation  \eqref{ds2Kb}.  The yellow triangle is a sketch of the emergent AdS$_2$ kinematic space $K_\p$, superimposed here on the picture of the bCFT background. %COMMENT ABOUT DIFFERENCE WITH dS. 
		Note that the kinematic space lightcone coordinates are just given by the bCFT lightcone coordinates, which is different from the situation in figure \ref{figK} for a CFT without boundary.   
		}
	\label{figbCFT}
\end{figure}

\begin{figure}[t]
	\begin{center}
		\includegraphics[]{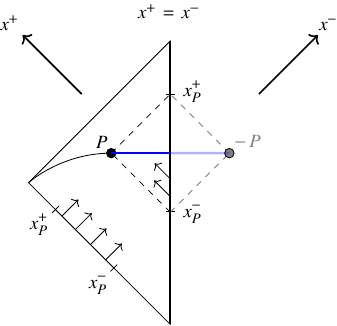} \qquad  
	\end{center}
	\caption{To write down the expressions for $S$ and $H_{mod}$ through the interval [$P$, boundary $x=0$], it is instructive to consider the doubled interval stretching from the point $P$ to its mirror image denoted $-P$. The formula \eqref{entbCFT} counts the entanglement contribution from right-moving degrees of freedom through the doubled interval, or the full entanglement through [$P$, boundary $x=0$].   
		We employ here the Penrose diagram representation of flat space with future and past null infinity at 45 degree angles. 
	} \label{figbCFTent}
\end{figure}

Now let us consider a CFT$_2$ in flat space $ds^2 = -dt^2 + dx^2$, $x \geq 0$ with a boundary at $x=0$ and large central charge $c$, that is in the vacuum state $|0\rangle_X$ with respect to the coordinate $X = X^+=X^-$. The thermal state e.g.~will have $X(x^\pm) = e^{2\pi x^\pm/\beta}$. 
We impose reflective boundary conditions 
\ali{
	T_{x^+x^+} = T_{x^-x^-}.  
}
Consider the interval that connects the point $P$ at $(x^+,x^-)$ to the boundary, as indicated in blue in figure \ref{figbCFT}.  
The presence of the boundary has the effect that the entanglement formula in \eqref{ent} now counts the entanglement through that interval from both right- and left-moving degrees of freedom: 
\ali{
	S(x^+,x^-) &= \frac{c}{12} \log \frac{(X^+(x^+) - X^-(x^-))^2}{\delta_x^2 \, {X^+}'(x^+) {X^-}'(x^-)} 
	= \frac{c}{12} \log \frac{(X(x^+) - X(x^-))^2}{\delta_x^2 \, {X}'(x^+) {X}'(x^-)} ,  \label{entbCFT}
}
with UV cutoff $\delta_x$ measured in $x^\pm$ coordinates. This is  illustrated in figure \ref{figbCFTent}. 
We omit here a possible constant contribution from the boundary entropy \cite{Cardy:2016fqc} and will not be concerned with the boundary dynamics of the theory, discussed in \cite{Callebaut:2018nlq}. 
The boundary CFT (bCFT) has a `boundary kinematic space' $K_\p$. 
Via the definition 
\ali{
	ds^2_{K_\p} &= -e^{\omega_x} dx^+ dx^- = -\left(\frac{2\ellK}{\delta_x}\right)^2 e^{-\frac{12}{c} S(x^+,x^-)} dx^+ dx^- , \label{ds2Kb} 
}
the entanglement determines the metric on $K_\p$ to be (a slicing of) the AdS$_2$ metric 
\ali{ 
	ds^2_{K_\p}	&= -\frac{4 \ellK^2 dX^+ dX^-}{(X^+-X^-)^2} = -\frac{4 \ellK^2 {X}'(x^+) {X}'(x^-) dx^+ dx^-}{(X(x^+)-X(x^-))^2}, \qquad X^+ = X(x^+), \, X^- = X(x^-).  	\nonumber
} 
Here we follow the discussion of the boundary kinematic space of a $d$-dimensional bCFT in \cite{Karch:2017fuh}, applied to $d=2$: for a CFT defined on $ds^2 = -dt^2 + dx^2$, $x \geq 0$ in the vacuum state $|0\rangle_X = |0\rangle_x$ (or uniformizing coordinates $X$ equal to the CFT coordinates $x$), 
the definition of kinematic space as the space of pairs of points \cite{Karch:2017fuh,Czech:2016xec} 
leads to the kinematic space metric $ds^2_{K_\p} = \frac{\ellK^2}{x^2} (-dt^2 + dx^2)$. A pair of points in this case refers to the point $P$ in figure \ref{figbCFTent} and its mirrored image across the boundary $x=0$. 
The kinematic space metric so obtained is the metric of AdS$_2$, rather than dS$_2$ in section \ref{LiouvilleKsection}. This explains our choice of sign in the definition \eqref{ds2Kb} of the boundary kinematic space metric in terms of the entanglement \eqref{entbCFT}.

The construction of $K_\p$ is summarized in figure \ref{figbCFT}. Compared to the dS kinematic space $K$, it are the lightcone coordinates of the CFT that become lightcone coordinates in $K_\p$, and the boundary of $K_\p$ that allows a natural association of just one point in $K_\p$ with one interval in the CFT is spacelike rather than timelike. 

Because of the identification between CFT and kinematic space lightcone coordinates, the vacuum entanglement Liouville stress tensor relates directly to the vacuum expectation value of the CFT stress tensor via 
\equ{ 
	\begin{aligned}  	
	T_{x^+x^+}^L &= \frac{1}{2\pi} \left( - \frac{6}{c} (\p_+ S)^2 - \p_+^2 S \right) = \frac{c}{24\pi} \{ X^+,x^+ \} = -\vev{T_{x^+x^+}}    \\ 
	T_{x^-x^-}^L &= \frac{1}{2\pi} \left( - \frac{6}{c} (\p_- S)^2 - \p_-^2 S \right) = \frac{c}{24\pi} \{ X^-,x^- \} = -\vev{T_{x^-x^-}}
	\end{aligned} \quad,    \label{TppL}  
}
compared to the analogue observation \eqref{TuuL} in section \ref{LiouvilleKsection}. 
We will comment on the interpretation of this relation at the end of section \ref{JacsectionK} (without being able however to elucidate the meaning of \eqref{TuuL}).

Similar to the entanglement formula, equation \eqref{HmodCFT} determines the full modular Hamiltonian (from both right- and left-moving degrees of freedom) through the interval that connects $P$ at $(x^+,x^-)$ to the boundary: 
\ali{
	H_{mod}(x^+,x^-) &= 2\pi \int_{x^-}^{x^+} ds \frac{\left( X(s)-X(x^-) \right) \left(X(x^+)-X(s)\right)}{\left(X(x^+)-X(x^-)\right) X'(s)}  \delta T_{x^+ x^+}(s).   \label{HmodbCFT}
}
Analogous to the discussion in section \ref{JTKsection}, we can write down the set of `boun\-dary kinematic space $K_\p$ identities'  	
\ali{
	\p_+ \p_- \left( \frac{12}{c} S \right) &= \frac{2}{\delta_x^2} e^{-\frac{12}{c} S}  \label{EOM1Kb} \\ 
	e^{-\frac{12}{c}S} \p_+ \left( e^{\frac{12}{c}S} \p_+ H_{mod} \right) &= 2 \pi \, \delta T_{x^+ x^+}(x^+) \label{EOM2Kb}  
	\\
	e^{-\frac{12}{c}S} \p_- \left( e^{\frac{12}{c}S} \p_- H_{mod} \right) &= 2 \pi \, \delta T_{x^- x^-}(x^-)  \label{EOM3Kb}    
	\\ 
	\p_+ \p_- H_{mod} &= -\frac{2}{\delta_x^2} e^{-\frac{12}{c}S} H_{mod}  \label{EOM4Kb} 
}
or in more transparent form, when expressed in terms of the natural metric $ds^2_{K_\p}$,  
\ali{
	R &=  -\Lambda \label{} \\ 
	\nabla_+ \p_+  H_{mod} &= 2\pi \, \delta T_{x^+x^+}(x^+)   \label{}\\
	\nabla_- \p_-  H_{mod} &= 2\pi \, \delta T_{x^-x^-}(x^-)  \label{}\\
	(\Box - \Lambda) H_{mod} &= 0 \label{}. 
}

By comparison to the JT equations of motion in AdS conformal gauge \eqref{EOM1}-\eqref{EOM4}, the following identifications between metric and entanglement and between dilaton and modular Hamiltonian can be made 
\ali{
	\omega &= - \frac{12}{c}S + 2 \log \frac{2 \ellK}{\delta}, \label{identifb0} \\ 
	\Phi &=  - 4G \, H_{mod} + \Phi_0. \label{identifb}
	}
The boundary kinematic space is thus governed by the JT action 	
\ali{
	I_{JT}[g,\Phi,\phi_m]  = \frac{1}{16 \pi G} \int d^2 \sigma 
	\sqrt{-g} \,  \Phi (R + \Lambda) \, \, + \, \, I_{m}[g,\phi_m]  \label{IJTKbaction}
	}
for a metric $ds^2 = g_\mn d\sigma^\mu d\sigma^\nu$ with lightcone coordinates $x^+$ and $x^-$, and a dilaton $\Phi$ (specified by equations \eqref{identifb0} and \eqref{identifb}), and conformal matter fields $\phi_m$.  The matter action $I_m$ of the JT theory of $K_\p$ is a bCFT on the AdS$_2$ kinematic space background, with stress tensor 
\ali{
	\delta T_{x^+x^+}(x^+) = T_{x^+x^+}^m(x^+). 
	}
We interpret this last equation as follows: the kinematic space $K_\p$ of a given bCFT is obtained by coupling that bCFT to AdS$_2$ JT gravity. Schematically, the kinematic space action \eqref{IJTKbaction} -- or in other words, the action that governs the entanglement dynamics of the bCFT -- can then be written as 
\ali{
	I_{\text{$K_\p$ of bCFT}} = \frac{1}{16 \pi G} \int d^2 \sigma 
	\sqrt{-g} \,  \Phi (R + \Lambda) \, \, + \, \, I_{\text{bCFT}} \, , 
	\label{Kbcoupling}
	}
where the metric and dilaton are respectively identified with the entanglement and modular Hamiltonian of the bCFT. 
This action expresses the identification of 
$H_{mod}$ as a collective mode of the CFT that 
obeys the $K_\p$ identities given above. 
Why was this interpretation not introduced in the discussion of the de Sitter kinematic space of a CFT (without boundary) in section \ref{JTKsection}? 
In that case, 
the JT matter stress tensor and the CFT stress tensor could not be equated.  Their relation was stated in equation \eqref{TmJTdS}. 
Indeed, it is the fact that the lightcone coordinates of the bCFT become the lightcone coordinates of the boundary kinematic space that allows to construct $K_\p$ from coupling the bCFT to JT gravity.

We take $\frac{\Phi_T}{4G} = - \vev{H_{mod}}_{\psi}$,  
or in different notation 
\ali{
	\frac{\delta \Phi}{4G} = - \delta \vev{H_{mod}},     \label{Kbconstruct}
}
as constructing principle for the semi-classical JT theory of $K_\p$.

\section{JT gravity and boundary kinematic space from maximal entanglement principle}  \label{Jacsection}

We have argued in the previous section that given a 2-dimensional CFT with boundary, its one-interval vacuum entanglement can be promoted to a field in kinematic space, the dynamics of which is described by a JT theory of 2-dimensional gravity coupled to the given boundary CFT. 
Similar elements occur in the maximal entanglement hypothesis put forward in \cite{Jacobson:2015hqa} (a related principle is discussed in \cite{Lloyd:2012du}), which is reviewed below. 
Paraphrased crudely, 
it states that imposing maximal entanglement in the vacuum state of a $d$-dimensional CFT amounts to coupling said CFT to gravity. 
Like other claims regarding the emergence of geometry from entanglement,  
it is based on reinterpreting a gravitational first law (in a theory of gravity) as a statement about entanglement in a CFT (without gravity). The former is given in equation \eqref{gravfirstlaw} and the latter in equation \eqref{reinterp} for the original $d$-dimensional argument, which was valid for $d>2$. 
We proceed to discuss a $d=2$ version of the argument in section \ref{JacsectionJT}. The strategy is to reinterpret the gravitational first law of AdS JT gravity given in equation \eqref{constraints} as a statement about entanglement, \eqref{JacJT}, in a 2-dimensional boundary CFT.  
We propose that the semi-classical JT equations of motion can indeed be obtained from a 2-dimensional version of Jacobson's maximal vacuum entanglement principle. 
Finally, we return to the boundary kinematic space context 
in section \ref{JacsectionK}. It is argued that the construction of boundary kinematic space amounts to imposing a maximal entanglement principle in the bCFT, effectively coupling it to semi-classical AdS JT gravity.

\subsection{Review of maximal entanglement principle} \label{Jacsectionreview}

Jacobson's maximal entanglement hypothesis \cite{Jacobson:2015hqa} is set in the context of quantum fields $\phi_m$ on a $d$-dimensional background geometry $g_\mn$. 
It states that the semi-classical Einstein equations $G_\mn + \Lambda g_\mn = 8 \pi G \langle T_\mn^m \rangle$ 
are equivalent to, and can be derived from, the statement that the vacuum state of the system $(g_\mn, \phi_m)$ has maximal entanglement, when $g_\mn$ is a maximally symmetric spacetime with cosmological constant $\Lambda$. The statement is most clear in the case of conformal matter $\phi_m$, which is the case we will restrict to in this paper.

Consider a 
spherical entangling surface $\mathcal B$ with causal domain of dependence $\mathcal D(\mathcal B)$ 
in a $d$-dimensional maximally symmetric background $g_\mn$ (gravity is `turned off', $G_d \ra 0$). 
The entanglement of quantum matter fields in $\mathcal B$ with the rest of the system typically has UV divergences, 
arising from infinitely many degrees of freedom at the boundary surface $\p \mathcal B$. The leading divergence scales with the area of $\p \mathcal B$: 
\ali{
	\Sjac = \# \frac{A(\p \mathcal B)}{\epsilon^{d-2}} + \text{  subleading divergences  } + S_{renormalized}.  \label{Sjac}
}  
One can now assume that unknown quantum gravity mechanisms impose finiteness of the entanglement $\Sjac$ by effectively 
imposing a UV cutoff equal to the Planck length 
$\epsilon = l_P$ (effectively `turning on' gravity, $G_d \neq 0$). 
As a consequence, the entanglement will contain an effective `geometrical part' $S_{UV}$ that represents the contribution from UV degrees of freedom, and a `matter part' $S$ for the IR degrees of freedom. 
The distinction between these contributions is ambiguous and renormalization scheme dependent, but the choice can be made to have metric variations only affect $S_{UV}$ and matter variations only affect $S_{IR}$.  
Under a simultaneous variation of $g_\mn$ and $\phi_m$, one can then write\footnote{
	In what follows we will simply write $\delta$ for all variations.
} 
\ali{
	\delta_{g,\phi_m} \Sjac = \delta_g S_{UV} + \delta_{\phi_m} S. \label{StotJac} 
} 

The difference between the surface area $A(\p \mathcal B)$ of a ball $\mathcal B$ in the perturbed geometry $g_\mn + \delta g_\mn$ and a ball $\mathcal B$ with the same volume in the unperturbed geometry $g_\mn$ is denoted $\delta A|_V$. A purely geometric relation 
relates the surface area difference to the Einstein tensor variation. The value of the cosmological constant $\Lambda$ 
of the maximally symmetric spacetime is not fixed by the argument but given at the onset\footnote{
	A similar remark applies in 
	\cite{Lloyd:2012du}.  
}. 
On the other hand, $\delta S$ can be written as a function of the variation of the matter stress tensor by making use of the first law of entanglement $\delta S = \delta \langle H_{mod} \rangle$. 
For $d>2$, these considerations allow to interpret the maximal entanglement condition 
\ali{
	\delta \Sjac|_V = 0 
}
as expressing the linearized semi-classical Einstein equations. By the use of Riemann Normal Coordinates these imply the non-linear equations of motion at the center of each ball, provided the radius of the ball is small enough compared to the curvature radius of the geometry, 
and thus the non-linear semi-classical Einstein equations $G_\mn + \Lambda g_\mn = 8 \pi G \langle T_\mn^m \rangle$ if the argument is to hold for all points in the geometry and in all frames.

\paragraph{First law derivation of maximal entanglement principle}

The equivalence of $\delta \Sjac|_V  = 0$ and the Einstein equations can alternatively be derived from a `first law of causal diamond mechanics' \cite{Jacobson:2015hqa,Bueno:2016gnv}.  
To obtain such a first law in a gravitational theory consisting of a metric and matter fields, one evaluates the relation \eqref{IWfirstlaw} 
for the region $\Xi = \mathcal B$ with conformal Killing vector $\xi$ satisfying $\xi|_{\p \mathcal B} = 0$:
\ali{
	\int_{\p\mathcal B} \delta Q_\xi -\delta H_\xi^g - \delta H_\xi^m = -\int_{\mathcal B} \delta C_\xi    \label{gravfirstlaw}
} 
for variations $\delta$ to a nearby solution.  
We imagine replacing all stress tensors in this identity by their (covariant) expectation value to obtain the semi-classical, linearized constraint equations on the right hand side. 
It is shown in \cite{Jacobson:2015hqa,Bueno:2016gnv} that $\delta H_\xi^g$ is proportional to the variation of the volume of the ball in the case of general relativity. Combined with the relation $\delta \vev{H_\xi^m} \sim \delta S$, which follows from the direct proportionality of the matter Hamiltonian with the modular Hamiltonian and the first law of entanglement, the left hand side can be written 
as 
\ali{
	\frac{\kappa}{2\pi} \delta S_{Wald}|_{V} + \frac{\kappa}{2\pi} \delta S = \frac{\kappa}{2\pi} \delta \Sjac|_{V}.   \label{reinterp}
} 	  
It follows immediately that $\delta \Sjac|_{V} = 0$ when the semi-classical, linearized constraint equations are satisfied, $\delta C_\xi (\vev{T_{ab}})  = 0$. 
The  `first law of causal diamond mechanics' is not to be interpreted as a physical process first law \cite{Gao:2001ut}, but as an equilibrium state first law \cite{Bueno:2016gnv}.

\subsection{Maximal entanglement principle applied to JT} \label{JacsectionJT}

The goal of this section is to reinterpret the JT first law 
as a maximal entanglement principle 
in  a 2-dimensional CFT on an AdS background. We follow the standard Iyer-Wald formalism in the discussion of the JT first law, and refer to section \ref{JTsectionMass} for a more detailed derivation of both the JT mass formula and first law.

\paragraph{JT first law} \label{firstlawsection}

\begin{figure}[h!]
	\begin{center} \includegraphics[]{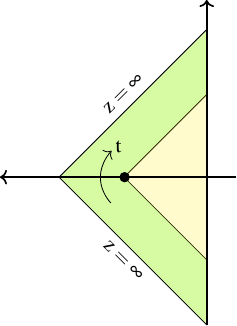}  \qquad \qquad  \includegraphics[]{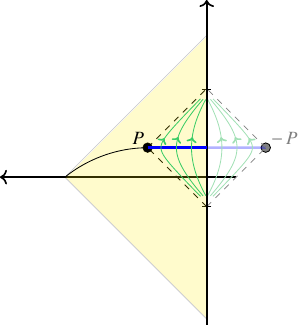}
	\end{center}
	\vspace{-2mm}
	\caption{\textit{Left:} The AdS$_2$-black hole solution of JT (yellow) and the Poincar\'e solution (largest triangular region), described by Poincar\'e covering coordinates $\tads$ and $\zads$. The Killing vector $\xi_{JT}$ associated with the Killing horizon of the black hole solution vanishes at the point $\{\tads = 0,\zads = \frac{1}{\sqrt{\mu}} \}$ labeled by the dot. The boundary is at $\zads=0$. The coordinate axes refer to global AdS$_2$ coordinates, see e.g.~\cite{Spradlin:1999bn,Maldacena:2016upp}. 
		\textit{Right:} The AdS$_2$-black hole solution %of figure \ref{figJTsol} 
		has a metric Killing vector $\xi_g$, with flow lines in green, that vanishes in $P$. The point $P$ at location $(X^+,X^-)$ or $(\tads=\tads_0,\zads=R)$ in Poincar\'e covering coordinates marks the boundary $\p \Sigma$ of the interval $\Sigma = \text{[$P$, boundary $\zads=0$]}$ (in blue). Imposing reflective boundary conditions at $\zads=0$, the interval can be effectively doubled to a region with diamond-shaped domain of dependence $\diamond$. 		
	}  
	\vspace{-3mm} \label{figJTsol}  % \label{HmodPfig}
\end{figure}

We discuss here the `first law' (or variational version of the JT mass formula derived in section \ref{JTsectionMass}) for the AdS$_2$ black hole solution of JT gravity. This solution is illustrated in figure \ref{figJTsol} and reviewed at some length in appendix \ref{JTreview}. 
The standard formalism for discussing gravitational first laws is that of Iyer and Wald \cite{Iyer:1994ys}, as reviewed in appendix \ref{IWappendix} where the (standard) notation is set.

JT theory has gravitational fields $\phi_g = \{g_\mn, \Phi\}$ and matter fields $\phi_m$. 
Evaluating equation \eqref{applytoJT} for the Killing vector $\xi_{JT}$ of the black hole solution 
gives rise to an expression of the form $\delta M = T \delta S_{bh} + \delta E_K$, with $E_K$ the Killing energy of matter fields, $M$ the mass, $T$ the temperature and $S_{bh}$ the Bekenstein-Hawking entropy of the black hole. In the standard interpretation, this formula compares to first order the thermodynamic quantities of a stationary black hole solution and another stationary black hole solution that is a linear perturbation away. Alternatively, in the `physical process' interpretation of \cite{Gao:2001ut}, it expresses the change in thermodynamic quantities as matter is thrown into an initially stationary black hole and it settles down into a final stationary state.

Now we similarly write a first law, not for the Killing vector $\xi_{JT}$ that vanishes at the real horizon of the black hole solution, but for the Killing vector $\xi_g$ that vanishes at the `horizon' at any point $P$:  
the JT solution in Poincar\'e covering coordinates  \eqref{AdS2metric}-\eqref{poincaresol} has a metric Killing vector $\xi_g$ that vanishes at the boundary $\p\diamond = \{P, -P \}$ of the doubled interval with diamond-shaped domain of dependence $\diamond$ presented in figure \ref{figJTsol} (right). By considering the diamond-shaped region we are assuming reflective boundary conditions at the AdS$_2$ boundary.
We evaluate the off-shell identity \eqref{IWfirstlaw} for the region $\Xi = \diamond$ and vector $\xi = \xi_g$ to obtain 
\ali{
	\delta H_{\xi_g}^g + \delta H_{\xi_g}^m &=  \int_{\p \diamond} \delta Q_{\xi_g} + \int_\diamond \delta C_{\xi_g}(E) \label{JTidentity}
}
(with Hamiltonian $H$, Noether charge $Q$ and $C(E)$ the constraint equations) for variations in the fields to a nearby 
solution. 

Let us discuss each term. The first term 
vanishes on account of the vector $\xi_g$ being a Killing vector of the metric, $\delta_{\xi_g} g_\mn = 0$, \emph{and} on account of the JT feature that the metric remains invariant, $\delta g_\mn = 0$.  
The gravitational part of the symplectic current $\omega_g $,  
which contains terms in $\delta_{\xi_g} \Phi \delta g_\mn$, $\delta \Phi \delta_{\xi_g} g_\mn$ and $\delta_{\xi_g} g_\mn \delta g_\mn$, then vanishes, even though the dilaton is not everywhere invariant under $\xi_g$. 
The second term is the matter Hamiltonian for 
flows along $\xi_g$, and is per definition \cite{Casini:2011kv,Hislop:1981uh} given by the modular Hamiltonian\footnote{\label{HmodPfootnote} Here we use the notation $H_{mod}^P$ for the modular Hamiltonian of the CFT on the AdS$_2$ background (with radius $\ell$), to distinguish it from the modular Hamiltonian $H_{mod}$ of the bCFT on a flat background considered in \eqref{HmodbCFT}. In section \ref{modhamsection} we will see however that they are in fact equal.}  $H_{mod}^P$ of the interval $\Sigma$ in figure \ref{figJTsol} (right). That is,  
$\delta H_{\xi_g}^m = - \frac{2}{\ell} \delta H_{mod}^P$. 
%$\delta H_{\xi_g}^m = -2 \delta E_{K,\triangleleft} = -2 \frac{\kappa}{2\pi} \delta H_{mod}^P$ in the notation of \eqref{canenergy}-\eqref{EKisHmod}.  
By %equation \eqref{SWaldxig}, 
standard Iyer-Wald formulation, the first term on the right hand side in 
\eqref{JTidentity} is equal to the Wald entropy $\frac{2}{\ell} \delta S_{Wald}$, which is a generalized Bekenstein-Hawking entropy for non-Einstein gravity theories. We thus find 
\ali{
	%-\frac{1}{\ell} \delta H_{mod}^P  &= \frac{1}{\ell} \delta S_{Wald} + \int_\Sigma \delta C_{\xi_g}(E). \\ %\label{constraints}
	-\delta H_{mod}^P  &= \delta S_{Wald} + \ell  \int_\Sigma \delta C_{\xi_g}(E). \label{constraints}	
}
It follows that on-shell,  
\ali{
	\delta H_{mod}^P = 	-\frac{\delta \Phi|_P}{4G},  \label{varmassform}
}
because the Wald entropy of a dilaton gravity theory is given by the dilaton evaluated at the `horizon', $S_{Wald} = \frac{\Phi}{4G}$ \cite{Nappi:1992as,Grumiller:2007ju}. 
%which we could have written immediately from equation \eqref{massformularesult}, but the IW notation will be useful in the next section. 

\paragraph{Reinterpreting JT first law as entanglement principle in bCFT on AdS background} 

It follows from equation \eqref{constraints} that the semi-classical, linearized constraint equations of motion of JT gravity can alternatively be expressed as a first law 
\ali{
	\delta S_{Wald} + \delta \vev{H_{mod}^P} = 0   
}
at any point $P$, or for any interval $\Sigma$. 
The first law of entanglement $\delta \vev{H_{mod}^P} = \delta S$ in the matter CFT of JT can subsequently be used.\footnote{ \label{deltaSfootnote} Here we could have employed  the notation $\delta S_P$ for the entanglement perturbations across the point $P$ of the CFT on the AdS$_2$ background, to distinguish it from the entanglement perturbations $\delta S$ of the bCFT on a flat background (as encountered in the context of section \ref{Kbsection}). However, based on the comment in footnote \ref{HmodPfootnote} which implies $\delta S = \delta S_P$, we can avoid this redundant notation.}  
The statement that the `total entanglement' is maximal  
\ali{
	\delta \Sjac = 0, \qquad \delta \Sjac = \delta S_{Wald} + \delta S    \label{JacJT}
}	
for all intervals $\Sigma$ in the matter bCFT and in all frames, then becomes equivalent to the semi-classical, linearized JT equations of motion (see  \eqref{JTEOM}) 
\ali{
	\delta \left( g_\mn \Box \Phi - \nabla_\mu \nabla_\nu \Phi - \frac{1}{2}g_\mn \Lambda \Phi \right) = 8 \pi G_2 \, \delta \langle T_\mn \rangle.  \label{linearizedJT}
}
As the variation on the left hand side only works on the dilaton ($\delta g_\mn = 0$), \eqref{linearizedJT} reduces to  
\equ{
	g_\mn \Box \delta \Phi - \nabla_\mu \nabla_\nu \delta \Phi - \frac{1}{2}g_\mn \Lambda \delta \Phi = 8 \pi G_2 \, \delta  \langle T_\mn \rangle. \label{linJT}
}
Making use of the linearity in the dilaton of the left hand side, these equations can be integrated directly to the full JT equations of motion 
\equ{
	g_\mn \Box \Phi - \nabla_\mu \nabla_\nu \Phi - \frac{1}{2}g_\mn \Lambda \Phi = 8 \pi G_2 \, \delta \langle T_\mn \rangle. \label{JTeqs}  
}
Compared to the general Jacobson argument there are some differences. One, we don't need to impose constant volume since $\delta H^g_{\xi_g} = 0$ in the JT theory. Second, the interval $\Sigma$ has arbitrary size, rather than being small. Indeed, we don't require the use of Riemann Normal Coordinates (and thus small radius of $\Sigma$) to integrate the linearized JT equations of motion to the full JT equations of motion, because of the linear (in the dilaton) nature of the JT model.

We formulate our conclusion as follows. Given a 2-dimensional CFT on a background with a boundary and negative cosmological constant $\Lambda$, imposing that the entanglement across any point $P$ in figure \ref{figJTsol} is maximal in the vacuum state, $\delta \Sjac = 0$, amounts to coupling the bCFT to semi-classical JT gravity.

Let us remark that we can also obtain the above as a 2-dimensional limit of the original derivation in \cite{Jacobson:2015hqa} (from the formula for $\delta A|_V$ rather than the first law argument), by making use of the techniques in \cite{Mann:1992ar}. That paper describes a $d \ra 2$ limit of Einstein gravity 
giving rise to dilaton gravity\footnote{
	We don't write the kinetic term in $\Phi$ that appears in the resulting action in \cite{Mann:1992ar}, because such an explicit kinetic term for the dilaton can 
	be tranformed away by a Weyl transformation (e.g.~\cite{Cadoni:1996bn}) 
	to obtain the form of the JT action as used  
	throughout the paper. Alternatively, \cite{Grumiller:2007wb} 
	obtains a Liouville dilaton gravity theory from a $d \ra 2$ limit of Einstein gravity.  
} 
\ali{
	\lim_{d\ra 2} \frac{1}{8 \pi G_d} G_\mn = \frac{1}{8\pi G_2} \left(g_\mn \Box \Phi - \nabla_\mu \nabla_\nu \Phi - \frac{1}{2}g_\mn \Lambda \Phi \right) 
}  
if the $d$-dimensional gravitational constant $G_d$ scales as 
\ali{
	\lim_{d\ra 2} 8 \pi G_d = \left(1 - \frac{d}{2} \right) 8 \pi G_2.  
}

\subsection{Maximal entanglement principle applied to boundary kinematic space}  \label{JacsectionK}

Because of the remarks in footnote \ref{HmodPfootnote} and  \ref{deltaSfootnote},  
we could also interpret \eqref{JacJT} as a maximal entanglement principle for a given 2-dimensional CFT on a background with a boundary and \emph{zero} cosmological constant $\Lambda$ (rather than negative $\Lambda$, as in the conclusion of the previous subsection). For the maximal entanglement principle to then be the expression of a gravitational first law of the type \eqref{JTidentity}, 
the type of gravity that the bCFT couples to has to have $\delta H^g_\xi = 0$, where $\xi$ is the kernel in the modular Hamiltonian \eqref{HmodbCFT} of the bCFT. As discussed in section \ref{firstlawsection}, this will be the case when both $\delta_\xi g_\mn = 0$ and $\delta g_\mn = 0$. The first condition imposes a hyperbolic metric -- in this interpretation $\Lambda$ is also emergent. The second condition is moreover true when the bCFT couples to JT dilaton gravity specifically, which has the property that the metric is always AdS$_2$.

It was argued in section \ref{Kbsection} that equation \eqref{Kbconstruct} can be taken as a constructing principle for the boundary kinematic space $K_\p$ of a given bCFT$_2$: the modular Hamiltonian $H_{mod}(x^+,x^-)$ defines a propagating field at the location $(x^+,x^-)$ in kinematic space via $\frac{\delta \Phi}{4G} = - \delta \vev{H_{mod}}$. This constructing principle takes the form of the 2-dimensional maximal entanglement principle \eqref{JacJT}, which expresses the 
coupling of the given bCFT$_2$ to JT gravity. 
We conclude that the boundary kinematic space of a bCFT$_2$ as defined in \cite{Karch:2017fuh} is obtained by coupling the bCFT$_2$ to AdS$_2$ JT gravity through a maximal entanglement principle.

The JT kinematic space theory is obtained from writing the kinematic space principle  \eqref{Kbconstruct} in the form \eqref{linJT} with $\delta \Phi = -4G \,  \delta \vev{H_{mod}}$, and integrating \eqref{linJT} to \eqref{JTeqs} while keeping the metric fixed, to find the $K_\p$ identities in  \eqref{EOM2Kb}-\eqref{EOM4Kb}.  Alternatively, the linearized equations \eqref{linJT} with $\delta \Phi = -4G \, \delta S$,  having used the first law of entanglement \eqref{FLE}, can be integrated to the Liouville equation \eqref{EOM1Kb} and the Liouville stress tensor in \eqref{TppL}:  
\ali{
	T_\mn^{L} = -\vev{ T_\mn } .  
}
This corresponds to integrating the linearized JT equations with the metric coordinates adjusted at each step to the uniformizing coordinates according to the Liouville field solution $S$, instead of keeping the metric coordinates fixed.

\section{JT model: entanglement considerations}  \label{JTsection} 

In this last section, we elaborate on some aspects of the JT model (with AdS$_2$ metric and conformal matter). It can be read as a stand-alone section, discussing respectively entanglement of the coupled CFT, the JT mass formula and the vacuum contribution to the dilaton. Nonetheless, 
we will conclude each subsection with comments relevant to kinematic space. 
Consider the Jackiw-Teitelboim model 
\ali{
	I[g,\Phi,\phi_m]  = \frac{1}{16 \pi G} \int d^2 \sigma \sqrt{-g} \Phi \left( R + \Lambda \right) \, \, + \, \, I_{m}[g,\phi_m]. \label{JTactionn} 
}
The matter part of the action $I_m$ describes a conformal field theory coupled to the metric $ds^2 = g_{\mn} d\sigma^\mu d\sigma^\nu$. 
This is a general Jackiw-Teitelboim theory 
with the %added 
assumption that the matter part of the action is independent of the dilaton. It is the action used in \cite{Engelsoy:2016xyb,Maldacena:2016upp,Jensen:2016pah} as bulk dual of a Schwarzian theory but we will not be concerned with that interpretation here (except for one related comment in section \ref{commentSchwarzian}). 
Instead we discuss the mass formula and first law that allow us to relate the modular Hamiltonian of conformal matter $\phi_m$ in an `entanglement wedge' with the value of the non-homogeneous contribution to the dilaton at the `entanglement wedge horizon', see equation \eqref{massformularesult}.

\subsection{Entanglement}   \label{JTsectionS}

Let us   
consider the CFT described by $I_m$ in the action  \eqref{JTactionn}. It is a 2-dimensional boundary CFT %(bCFT$_2$) 
coupled to the AdS$_2$ metric given in \eqref{AdS2metric} or \eqref{AdS2metricbhcoord}. 
The boundary of the metric is at $X^+=X^-$ in Poincar\'e coordinates or at $x^+=x^-$ in black hole coordinates. 

We consider the CFT to be in the vacuum state $|0\rangle_X$. When the parameter $\mu$ is non-zero in the JT solution, this corresponds through \eqref{Xxtransf} with a thermal state with respect to the black hole coordinate system. 
We will work in the covering coordinate system $X^\pm$ for the remainder of this section for the purpose of notational simplicity. 

The presence of the boundary allows us to associate an entanglement entropy and modular Hamiltonian with a \emph{point} $P$, which we will  %henceforth 
refer to in this section as $S_P$ and $H_{mod}^P$. We repeat that these are associated with the matter CFT on the AdS$_2$ geometry $ds^2 = -e^{\omega_X} dX^+ dX^-$ of the JT solution.  
Consider a point $P$ with coordinates $(X^+,X^-)$ and a Cauchy slice through $P$ that is divided in an `inside-$P$' and `outside-$P$' region, as shown in figure \ref{figJTsol} (right). We want to write down the expression for the entanglement $S_P$ across $P$ (see also \cite{Spradlin:1999bn}). We will follow \cite{Fiola:1994ir} in order to do so, where the following formula\footnote{
	This formula is derived by applying the standard formula for the vacuum one-interval entanglement in flat space $ds^2 = -dx_i^+ dx_i^-$ (with the index $i$ referring to inertial), but for a vacuum state $|0\rangle_X$ that is defined with respect to coordinates $X^\pm(x_i^\pm)$, in terms of which the metric takes the form $ds^2 = -e^{\omega_X} dX^+ dX^-$. Then $S_P = \frac{c}{6} \log \frac{L_X}{\delta_X} = \frac{c}{6} \log \frac{\delta_i}{\delta_X} + \frac{c}{6} \log \frac{L_X}{\delta_i} = \frac{c}{12} \omega_X + \frac{c}{6} \log \frac{L_X}{\delta_i}$, where $L_X$ denotes the length of the interval measured in $X$ coordinates, $\delta_X$ the UV cutoff measured in $X$ coordinates and $\delta_i$ the UV cutoff in $x_i$ coordinates. The resulting expression is then reinterpreted to give the formula in curved spacetime.  
} 
is given for the entanglement of an interval of length $(X^+-X^-)$ in a curved 2-dimensional background $ds^2 = -e^{\omega_X} dX^+ dX^-$: 
\ali{
	S_P = \frac{c}{12} \omega_X + \frac{c}{6} \log \frac{X^+-X^-}{\delta_i}. 
} 
Here  $\delta_i$ is a UV cutoff measured in local inertial coordinates at $P$. 
This expression includes the contribution from both right-moving and left-moving degrees of freedom, thanks to the reflective boundary conditions. 
In the metric under consideration \eqref{AdS2metric}, the conformal factor is 
\ali{
	\omega_X = 2 \log \frac{2 \ellads}{X^+-X^-}
} 
so that after substitution we find 
\ali{
	S_P = \frac{c}{6} \log \frac{2 \ellads}{\delta_i}. 
} 
That is, we find that the entanglement for the interval $\Sigma = [P,$ boundary $z=0]$ takes the form of a Rindler entropy \cite{Fiola:1994ir} with an IR cutoff that is given by the AdS radius. 
It follows that  
\ali{
	\omega = -\frac{12}{c} S + 2 \log \frac{2 \ellads}{\delta_i}  \label{JTentresult} \\
	\omega  = -\frac{12}{c} S + \frac{12}{c} S_P   \label{JTentresultb}
}
where we have used the notation $S$ for the `local' entanglement $\frac{c}{6} \log \frac{X^+-X^-}{\delta_i}$.

The local vacuum entanglement $S$ of the bCFT on AdS$_2$ maps to the conformal mode $\omega$ according to equation \eqref{JTentresult}. The JT equation of motion \eqref{EOM1} for the conformal mode thus imposes $S$ to satisfy the Liouville equation \eqref{EOM1Kb}. 

It further follows from equation \eqref{JTentresultb} that $\delta S = \delta S_P$ (this is the result referred to in footnote \ref{deltaSfootnote}) if we make use of $\delta \omega = 0$, i.e.~using once more that the background metric is unchanged under the variation to a nearby stationary JT solution.

\paragraph{Relevance to kinematic space} 

Since the expression for $S$ above matches the one for the entanglement \eqref{entbCFT} across the point $P$ in a bCFT on a flat background, the result in equation \eqref{JTentresult} %therefore 
explains why the vacuum entanglement of a boundary CFT satisfies a Liouville equation, i.e.~the first of four JT equations of motion, which was the starting point for the construction of boundary kinematic space. By extension, it gives immediate intuition for the perhaps arbitrary-looking identification in \eqref{Sofomega} of the vacuum entanglement in a CFT (without boundary) with the conformal mode of its kinematic space, which served as the definition of the kinematic space metric.  

Furthermore, the equality of $\delta S$ and $\delta S_P$ allowed us to reinterpret the JT first law as an entanglement principle in a bCFT on a \emph{flat} background in section \ref{JacsectionK}.

\subsection{Modular Hamiltonian}  \label{modhamsection}

To write down the formula for the modular Hamiltonian $H_{mod}^P$ for the same set-up, 
we need to discuss the `thermodynamics' associated with the point $P$, 
with which we can associate a Killing `horizon' by considering the Killing vector $\xi_g$ that vanishes in $P$. The flow lines of $\xi_g$ are shown in figure \ref{figJTsol} (right).  
Indeed the modular Hamiltonian will be determined by the Killing energy along those flow lines. 

The Killing vector of the AdS$_2$-Poincar\'e metric \eqref{AdS2metric} that acts within the triangular domain of dependence $\triangleleft$ of the interval depicted in figure \ref{figJTsol} (right) 
is given by 
\ali{
	\xi_g =  -\frac{\pi (-R^2 + (\tads-\tads_0)^2 + \zads^2)}{R \ellads} \, \p_{\tads} - \frac{2\pi \, (\tads-\tads_0) \, \zads}{R \ellads} \, \p_\zads  \label{xigAdS2}.
} 
It vanishes at the point $P$ with coordinates $\zads=R, \tads=\tads_0$. 
The subscript $g$ emphasizes $\xi_g$ is a Killing vector of the metric. The dilaton transforms non-trivially under it.  
We could introduce black hole coordinates $x^\pm$ whose full range cover only the domain of dependence $\triangleleft$: they are related to the Poincar\'e covering coordinates $X^\pm = \tads \pm \zads$ via $X^\pm = R \, \tanh \frac{x^\pm}{R} + \tads_0$. In terms of these coordinates the above Killing vector is just a black hole time translation $\xi_g = 2 \pi R \p_\tau$.

The surface gravity\footnote{
	We will drop quotes on the `thermodynamic' quantities associated with $P$ from now on.
} 
of $P$ is 
\ali{
	\kappa &= \left. \sqrt{\left|\frac{1}{2} \nabla^\mu \xi_g^\nu \nabla_\mu \xi^g_\nu \right|} \,  \right|_P = \left. \frac{\pi(R^2-(\tads-\tads_0)^2+\zads^2)}{\ellads R \zads} \right|_{\tads=\tads_0, \zads=R} = \frac{2\pi}{\ellads} 	\label{kappaxig}
}
and its temperature 
\ali{
	T = \frac{\kappa}{2\pi} = \frac{1}{\ell}.  \label{tempxig}
}
The Wald entropy, following the definition in terms of the Noether charge $Q(\xi_g)$ of \cite{Iyer:1994ys,Gegenberg:1994pv}, 
becomes  
\ali{
	S_{Wald} &= \frac{2\pi}{\kappa} \int_P Q(\xi_g) = \frac{2\pi}{\kappa} \frac{1}{16 \pi G} \left. \left(\epsilon_\mn \Phi \nabla^\mu \xi_g^\nu + 2 \epsilon_\mn \xi_g^\mu \nabla^\nu \Phi \right) \right|_P  = \frac{\Phi|_P }{4 G}  \label{SWaldxig}
}
because $\epsilon_{\mn} \nabla^\mu \xi_{g}^\nu|_P =  2 \kappa$ and $\epsilon_\mn \xi_g^\mu \nabla^\nu \Phi|_P = 0$ for any $P$ different from the horizon of the background.

The spacelike interval $\Sigma$ 
at $\tads=\tads_0$, with induced metric $h_{\zads\zads} = \ellads^2/\zads^2$, has a normal $n_\tads = - 1/\sqrt{|g^{\tads\tads}|} = - \ellads/\zads$ and a directed surface element $d\Sigma_{\tads} = - n_{\tads} \sqrt{|h|} d\zads = \ellads^2 d\zads / \zads^2$,  
consistent with  
$d\Sigma_\mu = \epsilon_{\mu \alpha} dx^\alpha$. 
Note that $d\Sigma^\tads = -d\zads$ is then past-directed per convention, leading us to define the Killing energy through $\Sigma$ with a minus sign:  
\ali{
	E_{K,\triangleleft}  &= -\int_\Sigma d\Sigma^\mu T_\mn^m \xi_{g}^\nu  
	= \frac{2\pi}{\ell} \int_0^R d\zads  \frac{R^2 - \zads^2}{2R} T_{00}^m(\tads_0,\zads).  
	\label{canenergy}
}
In terms of the lightcone coordinates  $(X^+ = t_0 + R, \, X^- = t_0-R)$ of the point $P$, we rewrite the energy to 
\ali{
	E_{K,\triangleleft}  &= \frac{2\pi}{\ell}  \int_{X^-}^{X^+} ds  \frac{(s-X^-)(X^+-s)}{X^+-X^-} T_{X^+X^+}^m(s=t_0+z) 
}
with $T_{00}^m = 2 T_{X^+X^+}^m$ because of reflective boundary conditions. 
The modular Hamiltonian of the CFT on the AdS$_2$ background is  
then given by \cite{Hislop:1981uh,Casini:2011kv}  
\ali{
	H_{mod}^P  &= \frac{2\pi}{\kappa} E_{K,\triangleleft} \label{EKisHmod}
} 
or 
\ali{
	H_{mod}^P = 2\pi \int_{X^-}^{X^+} ds \frac{(s-X^-)(X^+-s)}{X^+-X^-} T_{X^+X^+}^m(s).  \label{HmodP}
}
From comparison with the dilaton solution \eqref{PhiT} we can make the observation that the dilaton at location $(X^+,X^-)$ is related to $H_{mod}^P$ at that point by  
\ali{
	\Phi_T = -4G \, H_{mod}^P.  \label{phiJTHmod}
}
In the upcoming subsection we discuss the JT mass formula in order to derive the above relation between $\Phi$ and $H_{mod}^P$. 
From this relation it immediately follows that $H_{mod}^P$ satisfies the JT dilaton equations of motion \eqref{EOM2}-\eqref{EOM4}.

\paragraph{Relevance to kinematic space}

The Killing vector $\xi_g$ of the conformally flat background AdS$_2$ is also a conformal Killing vector $\xi$ of 2-dimensional flat space. As a result, $H_{mod}^P$ of the AdS$_2$ bCFT and $H_{mod}$ of the flat space bCFT, discussed in section \ref{Kbsection}, are given by the same formula, \eqref{HmodP} and \eqref{HmodbCFT} respectively.  
%Combined with equation \eqref{JTentresult}, it follows from the conclusion of the last paragraph that $H_{mod}$ and $S$ 
It then follows that $H_{mod}$ of the flat space bCFT %$_2$ 
should also satisfy JT equations of motion, as we indeed observed they do in \eqref{EOM1Kb}-\eqref{EOM4Kb}, interpreted there as kinematic space identities.  This reasoning implicitly equated the stress tensors of both bCFT's in equating their modular Hamiltonians, consistent with the interpretation of boundary kinematic space in \eqref{Kbcoupling} as the coupling of the bCFT to AdS$_2$ JT gravity.  %This already suggests the boundary kinematic space can be obtained from coupling the given bCFT to AdS$_2$ JT gravity. 

\subsection{Mass formula}  \label{JTsectionMass}

We are still considering the AdS$_2$-black hole solution of JT. Its metric has a horizon at 
$X^\pm = \pm \frac{1}{\sqrt{\mu}}$, 
and with that horizon we can associate a mass formula. 
For this purpose, we write down the 
Killing vector of the solution  
\ali{
	\xi_{JT}^\mu = \ellads \, \epsilon^\mn \nabla_\nu \Phi.  \label{ximuJT}
}  
Indeed, the dilaton is automatically constant along the Killing vector lines $\delta_{\xi_{JT}} \Phi = \xi^\mu_{JT} \nabla_\mu \Phi = 0$ 
and so is the metric $\delta_{\xi_{JT}} g_\mn = \nabla_\nu \xi_\mu^{JT} + \nabla_\mu \xi_\nu^{JT} = 0$.  

In 2 dimensions, a divergenceless current is always dual to the gradient of a scalar $M$ through $J_\mu = \epsilon_\mu^{\phantom{\mu}\nu} \nabla_\nu M$. For the JT black hole solution in absence of matter, the current $J_\mu = T_{\mn} \, \xi_{JT}^\nu = T_{\mn}^\Phi \, \xi_{JT}^\nu$, with $T_\mn^\Phi$ given in \eqref{TmnPhi}, is conserved by the definition of the Killing vector $\xi_{JT}$ in \eqref{ximuJT}. The corresponding mass function reads \cite{Mann:1992yv} 
\ali{
	M = -\frac{\ell}{16 \pi G} \left((\nabla \Phi)^2 - \frac{1}{\ell^2} \Phi^2 \right). 
}
It is constant on-shell, $\p_\alpha M = \epsilon_{\mu\alpha} T^\mn \xi_\nu^{JT}$, 
evaluating to $\frac{a^2 \mu}{16 \pi G \ell}$. 
This leads to a mass formula of the form $M_\infty = M_h$ or $2 M = T S_{bh}$ relating the mass, 
temperature and Bekenstein-Hawking entropy of the black hole solution \cite{Grumiller:2007ju}.

A stationary JT black hole solution in the presence of matter will in general have a Killing vector $\xi$ that is not equal to the one of the homogeneous solution $\xi_{JT}$. 
The Killing vector $\xi$ would equal $\xi_{JT}$ only when $T_\mn^m \sim g_\mn$, 
which is the case when the matter action $I_m$ is of the form $I_m = \int \sqrt{g} \, L_m(\phi_m, \Phi)$ (while we rather make the assumption that $I_m$ is independent of the dilaton, not the metric). But even for $\xi$ different from $\xi_{JT}$, the current $J_\mu = T_{\mn} \, \xi^\nu$ will be conserved as long as $\xi$ is the Killing vector of the metric\footnote{
	Note that $\xi$ need not be a Killing vector of the dilaton but only of the metric for this argument. 
}
and the energy-momentum tensor is conserved. 
The latter follows from diffeomorphism invariance of the gravitational theory. Because both the gravity part and the matter part of the action are separately diffeomorphism invariant 
we have $\nabla_\mu T^\mn = \nabla_\mu T^\mn_\Phi = \nabla_\mu T^\mn_m = 0$ and the divergenceless current $J_\mu = T_{\mn}^\Phi \, \xi^\nu$ allows the definition of a mass function (e.g.~\cite{McGuigan:1991qp}), via $\p_\alpha M = \epsilon_{\mu\alpha} T_\Phi^\mn \xi_\nu$, that is not constant on-shell, $\p_\alpha M = -\epsilon_{\mu\alpha} T_m^\mn \xi_\nu$. Integrating the last equation over the outside-horizon region gives rise to a mass formula 
with an extra contribution $E_K$ being the Killing energy of a matter fluid surrounding the black hole.

For the JT Poincar\'e solution with matter \emph{or} the JT black hole solution with matter in Poincar\'e covering coordinates one can consider the mass formula associated with the vector $\xi_g$, a  Killing vector of the metric (but not of the dilaton) that vanishes at the point $P$ at the position $(\tads=\tads_0, \zads = R)$ (see figure \ref{figJTsol}). 
In direct 
analogy with the preceding discussion, the conservation of the current $J_\mu = T_\mn^\Phi \, \xi_g^\nu$, with $T_\mn^\Phi$ defined in \eqref{TmnPhi} and $\xi_g$ in \eqref{xigAdS2}, 
determines an associated mass function through $\p_\alpha M_g = \epsilon^{\mu}_{\phantom{\mu}\alpha} T_\mn^\Phi \, \xi^\nu_g$. We find 
\ali{
	M_g = - \frac{1}{8\pi G} \frac{\pi}{\ellads R} 
	\left(\frac{R^2-(\tads-\tads_0)^2+\zads^2}{\zads}\Phi + (R^2-(\tads-\tads_0)^2-\zads^2) \p_\zads \Phi - 2 \, (\tads-\tads_0) \, \zads \, \p_\tads \Phi \right). \nonumber %  \label{Mg}
} 
Since $M_g$ is linear in $\Phi$, we have $M_g(\Phi) = M_g(\Phi_0) + M_g(\Phi_T)$ for which we will use the notation $M_g = M_g^0 + M_g^T$. On the dilaton solution \eqref{poincaresol} of the homogeneous equations $T_\mn^\Phi=0$, per definition $M_g$ is constant. Its constant value is related to the value of the dilaton at $P$ by $M_g^0 = -\Phi_{0}|_P/4G \ellads$. In the presence of matter however, $T_\mn^\Phi = -T_\mn^m$ and $M_g^T$ instead of being constant depends on $\zads$ (and $\tads$). 
When evaluated at the point $P$, $M_g^T$ does take the value of the dilaton $M_g^T = -\Phi_{T}|_P/4G \ellads$. 
We also wish to evaluate $M_g^T$ at the boundary $\zads \ra 0$ of $\Sigma$. This requires us to substitute the solution \eqref{PhiT} for $\Phi_T(X^+,X^-)$ written as a function of $\tads$ and $\zads$ using $X^\pm = \tads \pm \zads$,  
\ali{
	\Phi_T(\tads=\tads_0, \zads) = - 8\pi G \int_\zads^{-\zads} ds \frac{s^2-\zads^2}{2\zads} \frac{T_{00}(s)}{2}. 
}
It follows that in the limit $\zads\ra 0$, $M_g^T(\tads=\tads_0,\zads) \ra 0$. 
A mass formula is then obtained from the integration of $\p_\alpha M_g dx^\alpha = \epsilon^{\mu}_{\phantom{\mu}\alpha} T_\mn^\Phi \, \xi^\nu_g dx^\alpha$ over $\Sigma$ and on-shell evaluation $T_\mn^\Phi = -T_\mn^m$:
\ali{
	\int_{\p \Sigma} M_g = 
	-\int_\Sigma T_\mn^m \xi_g^\nu d\Sigma^\mu.  
}
The right hand side is the canonical energy $E_{K,\triangleleft}$  through $\Sigma$ as defined in \eqref{canenergy}. Further making use of equations \eqref{EKisHmod} and \eqref{tempxig}, the right hand side is then given by $H_{mod}^P/\ellads$.  The left hand side reduces to (leaving the evaluation at $\tads=\tads_0$ implicit) 
\ali{
	M_g(R) - M_g(0) &= M_g^0(R) + M_g^T(R) - M_g^0(0)  - M_g^T(0) \\
	&= M_g^T(R) = -\frac{\Phi_T|_P}{4G\ellads} \, , 
}
where in the second line we made use of the constancy of $M_0$ as well as the fact that $M_g^T(\zads \ra 0)$ vanishes. We finally are left with 
\ali{
	-\frac{\Phi_T|_P}{4G} = H_{mod}^P \, ,   \label{massformularesult}
}
meaning we have succeeded in understanding the origin of the observation \eqref{phiJTHmod} from 
the JT mass formula for the Killing vector $\xi_g$. 
We have shown that, schematically,  
that mass formula takes the form $M_\infty - M_h = E_K$ or $M_\infty = T S_{Wald} + E_K$. Subtracting from it the vacuum mass formula, and 
making use of the JT theory feature that all back-reaction is carried by the dilaton rather than the metric,   
gives $T \Delta S_{Wald} + E_K = 0$, with $E_K = H_{mod}^P/\ellads$ per definition vacuum-subtracted and $T \Delta S_{Wald} = 
\frac{\Phi_T}{4G \ellads}$, resulting in \eqref{massformularesult}.

\paragraph{Relevance to kinematic space}

The variational version of the JT mass formula or JT first law for the AdS$_2$ black hole solution was used in sections \ref{firstlawsection} and \ref{JacsectionK}, see in particular equation \eqref{varmassform}, to argue for a maximal entanglement principle interpretation of boundary kinematic space.

\subsection{Interpretation of $\Phi_0$}  \label{Phi0section}

Having established in \eqref{massformularesult} a relation between entanglement properties of the matter CFT of the JT black hole solution and the stress tensor dependent part of the dilaton, 
we investigate in this section the interpretation of the vacuum part $\Phi_0$ of the dilaton solution. Can $\Phi_0$ also be linked to entanglement of the matter CFT?  

In a coordinate system where the vacuum JT solution is static, 
the dilaton solution of the equations of motion \eqref{EOM4} and  \eqref{EOM2}-\eqref{EOM3}, $-e^\omega \p_\pm (e^{-\omega} \p_\pm \Phi) = 0$, can be written as\footnote{
	The coordinate $z$ is here e.g.~$\frac{X^+-X^-}{2}=\zads$ in the Poincar\'e solution \eqref{AdS2metric}-\eqref{poincaresol} (with $\mu=0$) or $\frac{x^+-x^-}{2}$ in the black hole solution \eqref{AdS2metricbhcoord}-\eqref{dilbhcoord}.   
} 
\ali{
	\Phi_0 = -\frac{a}{2 \ellads^2} \int^z dz' e^{\omega(z')}  ,
} 
where by equation of motion \eqref{EOM1} we have $e^\omega = \frac{1}{\Lambda} \p_z^2 \omega$, so that 
\ali{
	\Phi_0 = -\frac{a}{4} \p_z \omega.  
}
Under the identification \eqref{JTentresult} between the local entanglement $S$ across a point $P$ in the JT solution background and its conformal mode $\omega$, it follows $\Phi_0$ can be written in terms of $S$ as 
\ali{
	\Phi_0 =  3 \frac{a}{c} \p_z S.   \label{Phi0int1}
} 
Writing this as 
\ali{
	\Phi_0 &= 
	3 \frac{a}{c}  \frac{S(z+ \epsilon) - S(z)}{\epsilon}  \label{misalignment}
}
explains the misalignment interpretation of $\Phi_0$ in \cite{Callebaut:2018nlq}: 
a small change in the location of the boundary results in a removal of entanglement in an amount equal to 
$\Phi_0$, 
\ali{
	\frac{\Phi_0}{4G_2} &= 
	S(z+ \epsilon) - S(z),   \qquad a = 4 G_2 \frac{c \epsilon}{3}
} 
if $a$ is related to the central charge $c$ via $a = 4 G_2 c \epsilon/3$ (or $C = c \epsilon / 12 \pi$ in the notation of \cite{Callebaut:2018nlq}).

The JT solution can be seen as the spherical dimensional reduction of an asymptotically AdS$_3$ parent theory \cite{Achucarro:1993fd,Grumiller:2002nm}, which has a 2-dimensional dual CFT with central charge $\tilde c$ (distinguishing $\tilde c$ here from the central charge $c$ of the 2-dimensional matter CFT in the JT bulk). We can then alternatively identify $\omega$ with the entanglement $S$ of an interval of that CFT through $\omega = -\frac{6}{\tilde c} S$ (up to a constant), allowing $\Phi_0$ to be interpreted as differential entropy 
\ali{
	\frac{\Phi_0}{4G_2} &=  
	S_{\mathit{diff}},   \qquad a = 4 G_2 \pi \frac{\tilde c \mathcal R}{3}  \label{Phi0Sdiff}
} 
where $\mathcal R$ is the radius of the conformal boundary 
of asymptotically AdS$_3$ and $S_{\mathit{diff}}$ is defined \cite{Balasubramanian:2013lsa} 
as 
\ali{
	S_{\mathit{diff}} = \left. \pi \, \p_\alpha S \, \right|_{\alpha \mathcal R = 2z}  \label{Sdiffdef}
}
in terms of the entanglement $S$ of an interval of length $2 z$ or 
angular size $2\alpha$. 
The relation between the dilaton and the differential entropy follows very naturally from the 3-dimensional `parent' picture and is illustrated in figure \ref{figSdiff}. We discuss it in some more detail below.

\subsubsection{Intuition from dimensional reduction}

\begin{figure} 
	\begin{center} \includegraphics[]{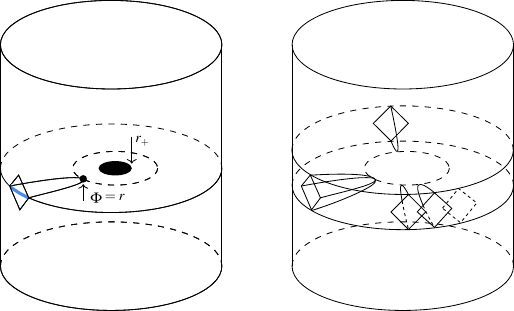} 
	\end{center}  
	\caption{\textit{Left:} Dilaton as differential entropy \eqref{Sdiffdef} from considering the parent asAdS$_3$ theory with possible horizon at $r=r_+$. This relates the dilaton with the entanglement of the interval in blue, different from the interpretation \eqref{Phi0int1} that relates the dilaton with the entanglement of the interval in blue in figure \ref{figJTsol} (right). \textit{Right:} Differential entropy measures `entanglement' of the strip of width $2 z$. 
	}
	\vspace{-3mm} \label{figSdiff}
\end{figure}

For a 3-dimensional metric that is separable and spherically symmetric, 
\ali{
	ds_3^2 = g_{3,\mn}(x) dx^\mu dx^\nu = g_{2,ij}(x^i) dx^i dx^j + \psi^2(x^i) d\phi^2 , \label{ds23D}
}
one has $\sqrt{g_3} = \sqrt{g_2} \psi$ and $R_3 = R_2 - 2 \frac{\Box \psi}{\psi}$, 
so that 
the 3-dimensional Einstein-Hilbert action $I_{EH}$ takes the form 
\ali{
	16 \pi	\, I_{EH} &= \frac{1}{G_3} \int d^3 x \sqrt{g} (R_3 + \Lambda) = \frac{2\pi}{G_3} \int d^2 x \sqrt{g_2} \psi (R_2 + \Lambda). \label{EHaction} 
}	
A solution 
of the form \eqref{ds23D} then directly gives rise to a solution of the 2-dimensional dilaton gravity action  \cite{Achucarro:1993fd} 
\ali{
	ds^2_2 =  g_{2,ij}(x^i) dx^i dx^j, \qquad  \psi(x^i),  
} 	
where the dilaton $\psi$ measures the radial coordinate in \eqref{ds23D}. For example, the BTZ solution 
\ali{
	ds_3^2 = -\left(\frac{r^2 - r_+^2}{\ell^2} \right) d\tau^2 + \left(\frac{r^2- r_+^2}{\ell^2} \right)^{-1} dr^2 + r^2 d\phi^2 
}
with horizon 
\ali{
	r_+ = \sqrt{\mu} \ellads \mathcal R \,\, , \quad \mu = \frac{2\pi}{\beta},  
}	
given in terms of the radius $\mathcal R$ of the conformal boundary 
of BTZ ($ds^2_3 \stackrel{r \ra \infty}{\ra} \frac{r^2}{\mathcal R^2} ds^2_{\text{conformal bdy}}$) 
and the inverse temperature $\beta$, 	
gives rise to the AdS$_2$-black hole solution 
\ali{
	ds_2^2 = -\left(\frac{r^2 - r_+^2}{\ell^2} \right) d\tau^2 + \left(\frac{r^2- r_+^2}{\ell^2} \right)^{-1} dr^2, \qquad \psi = r 
}
with the same parameter $\mu$. 
The dilaton $\psi$ 
in \eqref{EHaction} will always appear in the dimensionless combination $\psi/G_3$. It is then useful 
to introduce a dimensionless dilaton $\Phi$ so that 
\ali{
	2\pi \frac{\psi}{G_3} = \frac{\Phi}{G_2}  \label{psiphi}
}
defines a type of effective, running gravitational constant 
$G_{2,\mathit{eff}}$.  
In terms of $\Phi$ the action \eqref{EHaction} takes the JT form, and the  AdS$_2$-black hole solution above matches the notation in  \eqref{AdS2metricbhcoord}-\eqref{dilbhcoord} (with $\mathcal R = \ellads$) if we define 
\ali{
	G_3 &= \frac{2\pi \ellads \mathcal R}{a} G_2 \label{G3ofG2} \\
	\Phi &= \frac{a}{\ellads \mathcal R} \psi  .  
}
Here the arbitrary length scale $a$ was introduced to extract a dimensionless dilaton from $\psi$ (it is related to the arbitrary mass scale $\lambda$ of \cite{Grumiller:2002nm} via $\lambda = \frac{a}{\ellads \mathcal R}$). Note also that $\psi = r_+ \coth(2\sqrt{\mu}\,z)$ 
matches the expression for the depth 
reached by a geodesic anchored to a BTZ boundary interval of length $2z$, as it should. 

It follows directly from \eqref{psiphi} that the dilaton  
\ali{
	\frac{\Phi}{4G_2} = \frac{A}{4G_3}
}
measures the area $A = 2\pi r$ of the `hole' of radius $r$ in the 3-dimensional background. 
It is argued in \cite{Balasubramanian:2013lsa} that, while the area of a Ryu-Takayanagi surface in the locally AdS$_3$ bulk is measured by one-interval entanglement $S$ of the dual CFT, the area of the hole in the bulk has to be measured as the envelope of a collection of Ryu-Takayanagi geodesics of fixed opening angle $\alpha$ as illustrated in figure \ref{figSdiff}. It is then the quantity $\pi \,  \p_\alpha S$ or `differential entropy' $S_{\mathit{diff}}$ that forms the CFT dual of the gravitational entropy $\frac{A}{4G_3}$ of the hole, 
\ali{
	S_{\mathit{diff}} = \frac{A}{4G_3}, 
}
and thus by the previous relation $\frac{\Phi}{4G_2}$. 
This establishes equation \eqref{Phi0Sdiff}. In particular, the relation in \eqref{Phi0Sdiff} between $a$ and the (effective) central charge $\tilde c$ of the dual CFT 
follows from the standard 3d/2d holographic dictionary entry \cite{Brown:1986nw} 
\ali{
	\tilde c = \frac{3 \ellads}{2G_3}
}  	
and equation \eqref{G3ofG2} for $G_3$ as a function of $G_2$. 
The same relation $a(\tilde c)$ appeared in \cite{Engelsoy:2016xyb} as a condition under which the Bekenstein-Hawking entropy of the JT solution 
takes the form of a Cardy formula. Indeed, this similarly follows from spherical dimensional reduction of the statement \cite{Strominger:1997eq} that the Bekenstein-Hawking entropy of BTZ 
maps to 
the Cardy formula for the entropy of a thermal CFT$_2$ counting the number of states with a given conformal dimension 
on the cylinder.

In the dual CFT, $S_{\mathit{diff}}$ has the interpretation of `entanglement' of the strip of width $2z$ in the time direction, with caveats for referring to it as an actual entanglement discussed in \cite{Balasubramanian:2013lsa,Czech:2014tva}. Let us remark here that  
equation \eqref{Phi0Sdiff} is then suggestive of the JT dilaton being holographically dual to the `entanglement' of a time-like interval of length $2z$. It is unclear how to interpret such an object in the 1-dimensional dual theory. 
In this context it is perhaps important to stress that the one-interval entanglement $S$ in \eqref{Phi0Sdiff}-\eqref{Sdiffdef} is the one for a small interval compared to the size of the system. This means that for e.g.~the BTZ background with conformal boundary the torus $\mathcal T(2\pi \mathcal R, \beta)$, it is given by 
\ali{
	S = \frac{\tilde c}{3} \log \left( \frac{\beta}{\epsilon \pi} \sinh \frac{2 \pi \alpha \mathcal R}{\beta} \right).  
}
This is 
the correct expression for the entanglement in the high temperature limit $\mathcal R/\beta \ra \infty$ or for small enough intervals. For intervals larger than a certain threshold known as the `entanglement shadow', $S$ measures the length of winding geodesics in the bulk or `entwinement' \cite{Balasubramanian:2014sra} in the dual CFT (rather than the length of the minimal geodesic in the bulk or entanglement in the dual CFT). 
The 1-dimensional theory dual to JT is obtained from the 
Liouville description of the 2-dimensional dual of asymptotically AdS$_3$ in the 
limit $\mathcal R/\beta \ra 0$ \cite{Mertens:2017mtv,Mertens:2018fds}, where $S$ becomes pure entwinement.

\paragraph{Relevance to kinematic space}   \label{commentSchwarzian}

%cMERA removing a small amount of entropy by shifting the boundary inward
The interpretation of the vacuum contribution to the dilaton in \eqref{misalignment} as the amount of entanglement removed as a result of a displacement of the boundary, is used in \cite{Callebaut:2018nlq} to obtain an entropic derivation of the Schwarzian theory that describes the dynamics of the boundary in JT gravity. The argument makes use of entanglement renormalization, and is reminiscent of the concept of cMERA \cite{Haegeman:2011uy}, short for continuous Multiscale Entanglement Renormalization Ansatz, with MERA a real-space renormalization group method in quantum many-body physics \cite{Vidal:2007hda}.  
The description of kinematic space as a JT theory makes  
the relation between cMERA and the Schwarzian theory  
particularly  
natural, given the kinematic space/MERA proposal put forward in \cite{Czech:2015qta}, which argues to consider %the quantum many-body physics 
MERA a discretization of kinematic space. 
It would be interesting to study the relation between cMERA and the Schwarzian theory further.

\section{Discussion}  \label{discussion}

We have discussed the JT dynamics of the kinematic space of 2-dimensional CFT's with or without a boundary. The corresponding kinematic space  respectively has an AdS or dS metric. 
The motivation for treating the `kinematic space identities' in \eqref{EOM1K}-\eqref{EOM4K} and \eqref{EOM1Kb}-\eqref{EOM4Kb} as equations of motion 
initially was the mere observation that they coincided with the JT equations of motion. Doing so results in the concept of treating entanglement itself as a dynamic field.   
The goal of the paper is to provide an interpretation for the resulting JT theory, which turned out to be more straightforward in the case of boundary kinematic space, without claiming a complete analysis of the construction or a full understanding of its generality. 
In section \ref{Jacsection} it was argued that the constructing principle \eqref{Kbconstruct} of boundary kinematic space $K_\p$ can be interpreted as a 2-dimensional version of Jacobson's maximal entanglement principle that couples the given bCFT to JT gravity on AdS. This discussion complements the results on the entanglement dynamics of a boundary CFT in \cite{Callebaut:2018nlq} by providing a kinematic space point of view.  
It remains less clear however if a similar statement can be made for the de Sitter kinematic space $K$ discussed in section \ref{JTKsection}.  
What can be repeated for de Sitter, with the same conclusions, is the JT gravity discussion in section \ref{JTsection} leading to equation  \eqref{massformularesult}, with the remark that $\xi_g$ being a spacelike rather than a timelike Killing vector renders the `thermodynamic quantities' of section \ref{modhamsection} less physical meaning.  

There is another context in which the coupling of a CFT to JT(-like) gravity appears, namely in the $T\bar T$ theory obtained by turning on a $T\bar T$ deformation of the CFT \cite{Smirnov:2016lqw,Dubovsky:2017cnj,Dubovsky:2018bmo}. It would be interesting to study any connection of the $T\bar T$ theory to this work. 

We would also like to understand better the relations between the different Liouville theories that appear in the AdS$_3$/CFT$_2$ context: the kinematic space Liouville theory of section \ref{LiouvilleKsection} and the Liouville theory describing the asymptotic dynamics of AdS$_3$, as well as the Liouville theory associated with complexity of \cite{Caputa:2017yrh}. 
Related questions are raised by the discussion of the interpretation of the JT dilaton  from a dimensional reduction from AdS$_3$ standpoint in section  \ref{Phi0section}. 

We leave these problems for future study.

\acknowledgments

I am very grateful to Herman Verlinde for many insightful discussions and collaboration on a closely related paper. I further would like to thank Thomas Mertens in particular, for many helpful comments and discussions, as well as Steve Carlip, Bartek Czech, Ted Jacobson, Finn Larsen, Aitor Lewkowycz, Seth Lloyd, Mark Mezei, Charles Rabideau, Antony Speranza, Joaquin Turiaci, and Zhenbin Yang. This work is supported by the Research Foundation-Flanders (FWO Vlaanderen).

\appendix

\section{Jackiw-Teitelboim gravity} \label{JTreview}

\paragraph{JT action and equations of motion}

The Jackiw-Teitelboim theory \cite{Jackiw:1984je,Teitelboim:1983ux} is the dilaton gravity theory 
\ali{
	I[g,\Phi,\phi_m]  = \frac{1}{16 \pi G} \int d^2 \sigma \sqrt{-g} \left(\Phi R - V(\Phi) \right) \, \, + \, \, I_{m}[g,\phi_m]  \label{JTaction}
}
with a linear potential 
\ali{ 	V(\Phi) = -\Lambda \Phi, } 
and a matter action $I_{m}(g,\phi_m)$ that describes a field theory coupled to the metric $ds^2 = g_{\mn} d\sigma^\mu d\sigma^\nu$. 
We assume that field theory to be conformal, and furthermore assume $I_m$ to be independent of the dilaton, such that variation with respect to the dilaton inforces the constant curvature equation 
\ali{
	R &= -\Lambda. 
}
Variation with respect to the metric gives 
\ali{
	T_\mn  &= T_{\mn}^\Phi + T_{\mn}^m = 0 
}
with 
\ali{
	T_{\mn}^\Phi & 
	= 	-\frac{1}{8\pi G} ( g_\mn \Box \Phi - \nabla_\mu \nabla_\nu \Phi - \frac{\Lambda}{2}g_\mn \Phi ) \label{TmnPhi} \\ 
	T_{\mn}^m &= -\frac{2}{\sqrt{-g}}\frac{\delta I_m}{\delta g^\mn}. 
}
The equation of motion 
\ali{ 
	g_\mn \Box \Phi - \nabla_\mu \nabla_\nu \Phi + \frac{1}{2}g_\mn V 	= 8\pi G T_{\mn}^m \label{JTEOM}
} 
can be split up in a traceless and a trace part 
\ali{
	\frac{1}{2}g_\mn \Box \Phi - \nabla_\mu \nabla_\nu \Phi &= 8\pi G  (T^m_{\mn} - g_\mn T_{\,\,\sigma}^{m \, \sigma}) \\ 
	(\Box - \Lambda) \Phi &= 8\pi G  T_{\,\,\sigma}^{m \, \sigma}.
}
For (classical) conformal matter, $T_{\,\,\sigma}^{m \, \sigma}=0$, and in `AdS' conformal gauge $ds^2 = -e^{\omega(x^+,x^-)}$ $dx^+ dx^-$ ($\Ra \Box = -4 e^{-\omega} \p_+ \p_-$, $R = 4 e^{-\omega} \p_+ \p_- \omega = -\Box \omega$), the JT equations of motion (EOM) read 
\ali{
	4 \p_+ \p_- \omega + \Lambda e^{\omega} &= 0 \qquad \qquad \quad  ( R = -\Lambda) \label{EOM1} \\ 
	-e^\omega \p_+ (e^{-\omega} \p_+ \Phi) = 	\p_+ \Phi \p_+ \omega - \p_+^2 \Phi &= 8\pi G T_{++}^m \label{EOM2}  \qquad (-\nabla_+^2 \Phi = 8\pi G T_{++}^m) \\
	-e^\omega \p_- (e^{-\omega} \p_- \Phi) = 	\p_- \Phi \p_- \omega - \p_-^2 \Phi &= 8\pi G T_{--}^m \label{EOM3}  \qquad  (-\nabla_-^2 \Phi = 8\pi G T_{--}^m)\\ 
	\p_+ \p_- \Phi + \frac{\Lambda}{4} e^\omega \Phi &= 0 \qquad\qquad \quad  (\Box \Phi - \Lambda \Phi = 0).  \label{EOM4} 
} 
If we change the sign of the potential to $V(\Phi) = \Lambda \Phi$ such that $R = \Lambda > 0$ is imposed, then in `dS' conformal gauge $ds^2 = e^{\omega(x^+,x^-)} dx^+ dx^-$ ($\Ra \Box = 4 e^{-\omega} \p_+ \p_-$, $R = -4 e^{-\omega} \p_+ \p_- \omega = -\Box \omega$), the JT EOM \eqref{EOM1}-\eqref{EOM4} are unchanged up to the sign of $\Lambda$:   
\ali{
	4 \p_+ \p_- \omega + \Lambda e^{\omega} &= 0 \qquad  \qquad \quad ( R = \Lambda) \label{EOM1dS} \\ 
	-e^\omega \p_+ (e^{-\omega} \p_+ \Phi) = 	\p_+ \Phi \p_+ \omega - \p_+^2 \Phi &= 8\pi G T_{++}^m \qquad (-\nabla_+^2 \Phi = 8\pi G T_{++}^m) \label{EOM2dS}\\
	-e^\omega \p_- (e^{-\omega} \p_- \Phi) = 	\p_- \Phi \p_- \omega - \p_-^2 \Phi &= 8\pi G T_{--}^m \qquad (-\nabla_-^2 \Phi = 8\pi G T_{--}^m) \label{EOM3dS}\\ 
	\p_+ \p_- \Phi + \frac{\Lambda}{4} e^\omega \Phi &= 0 \qquad  \qquad \quad (\Box \Phi + \Lambda \Phi = 0).\label{EOM4dS} 
}

Semi-classically, the trace anomaly $T_{\,\,\sigma}^{m \, \sigma}= \frac{c}{24\pi} R$ is taken into account by considering the effective action $I_{\mathit{eff}} = I_{JT} + I_{Pol}$, with the Polyakov action  \cite{Almheiri:2014cka,Engelsoy:2016xyb}
\ali{
	I_{Pol} = -\frac{c}{96 \pi} \int d^2x \sqrt{-g} \, R \frac{1}{\Box} R  	\label{IPol}
}
with $c$ large. 	
The semi-classical JT EOM then become (in AdS conformal gauge) 	
\ali{
	4 \p_+ \p_- \omega + \Lambda e^{\omega} &= 0 \qquad    && ( R = -\Lambda)  \\ 
	\p_+ \vev \Phi \p_+ \omega - \p_+^2 \vev \Phi &= 8\pi G \vev{T_{++}^m} && (-\nabla_+^2 \vev \Phi = 8\pi G \vev{T_{++}^m}) \\
	\p_- \vev \Phi \p_- \omega - \p_-^2 \vev \Phi &= 8\pi G \vev{T_{--}^m}  && (-\nabla_-^2 \vev \Phi = 8\pi G \vev{T_{--}^m})\\ 
	\p_+ \p_- \vev \Phi + \frac{\Lambda}{4} e^\omega \vev \Phi &= 8\pi G \vev{T_{+-}^m} &&  (\Box \vev \Phi - \Lambda \vev \Phi = 8 \pi G \vev{T_{\,\,\sigma}^{m \, \sigma}}), 
} 	
with covariant stress tensor components $\vev{T_{\mn}^m}$, and in particular  $\vev{T_{+-}^m} = - \frac{c}{24\pi} \p_+ \p_- \omega$ 
so that, upon use of the Liouville equation $R = -\Lambda$, the last EOM reads 	
\ali{
	\p_+ \p_- \vev \Phi + \frac{\Lambda}{4} e^\omega \left(\vev \Phi - \frac{c G}{3} \right)  &=  0.  
} 
It follows that the solution $\vev \Phi$ is related to a solution  $\tilde \Phi$ of the classical EOM by $\vev \Phi = \tilde \Phi + \frac{c G}{3}$ (and $T_\mn^m$ replaced by $\vev{T_\mn^m}$). Assuming the constant shift can be absorbed in the vacuum contribution $\vev{\Phi_0} = \tilde \Phi_0 + \frac{cG}{3}$, the vacuum-subtracted semi-classical dilaton $\vev{\Phi_T}$ obeys     
\ali{
	-\nabla_\pm^2 \vev{\Phi_T} &= 8\pi G \vev{T_{\pm\pm}^m} \\
	\Box \vev{\Phi_T} - \Lambda \vev{\Phi_T} &= 0
} 
as if it were a classical dilaton $\tilde \Phi_T$, with $\vev{T_{\pm\pm}^m}$ the vacuum-subtracted, covariant stress tensor expectation value.

\paragraph{Solutions of JT} 

The general solution of the \emph{homogeneous} JT EOM, \eqref{JTEOM} with $T_\mn^m = 0$, is given by an AdS$_2$-black hole metric and a dilaton profile, which in Poincar\'e covering coordinates $X^\pm = \tads \pm \zads$ reads
\ali{
	ds^2 &=  \frac{\ellads^2}{\zads^2} (-d\tads^2 + d\zads^2) = - \frac{4 \ell^2 dX^+ dX^-}{(X^+-X^-)^2} 
	\label{AdS2metric}
	\\ 
	\Phi_0 &= a \frac{1 - \mu (\tads^2 - \zads^2)}{2\zads} = a \frac{1 - \mu \, X^+ X^-}{X^+ - X^-}.   \label{poincaresol}	
} 	
Here $a$ and $\mu$ are integration constants with dimension of length and one over length squared respectively, and we use the notation $\Phi_0$ for the dilaton to indicate it is a vacuum solution. 
The geometry, illustrated in figure \ref{figJTsol} (left), spans a triangular region with a boundary at $X^+ = X^- = \tads$.  
The quantity $\mu$ is related to the energy of the black hole solution and vanishes in the Poincar\'e solution. 

The solution can alternatively be written in black hole coordinates $(\tau,r)$ or $x^\pm = \frac{\tau}{2} \pm z(r)$ that cover the black hole triangle and are natural from a dimensional reduction viewpoint \cite{Achucarro:1993fd},  
\ali{
	ds^2 &= -\left(\frac{r^2 - \ell^4 \mu}{\ell^2} \right) d\tau^2 +  \frac{dr^2}{\left(\frac{r^2- \ell^4 \mu}{\ell^2} \right)} = -4 \ell^2 \mu  \, \csch^2 \left(\sqrt{\mu} (x^+-x^-)\right)  dx^+ dx^-  \label{AdS2metricbhcoord} \\ 
	\Phi_0 &= \frac{a}{\ell^2} r = a \sqrt{\mu} \, \coth(\sqrt{\mu}(x^+-x^-)).   \label{dilbhcoord}
}   	
In these coordinates, the Killing horizon of the solution is at $z \ra \infty$ or $r = \ell^2 \sqrt{\mu}$, where the dilaton takes the value \ali{ 
	\Phi_{0,h} = a \sqrt{\mu}.
}  
The transformation to the Poincar\'e covering coordinates is given by 
\ali{
	X^\pm(x^\pm) = \frac{1}{\sqrt{\mu}} \tanh(\sqrt{\mu} \, x^\pm).  \label{Xxtransf} 
}

The general solution of the \emph{inhomogeneous} JT EOM \eqref{JTEOM} still has the same constant curvature metric. 
The dilaton however receives a stress tensor dependent contribution, which we will denote $\Phi_T$: 
\ali{
	\Phi &= \Phi_0 + \Phi_T \\ 
	\Phi_{T} 
	&= -8 \pi G \int_{X^+}^{X^-} ds \frac{(s-X^+)(s-X^-)}{X^+-X^-} T_{X^+ X^+}^m(s). \label{PhiT} 	
}
To infer this form of the solution from its more general form in terms of integration functions $I_\pm$ in \cite{Almheiri:2014cka} and \cite{Engelsoy:2016xyb} requires two remarks. First, we have imposed reflective boundary conditions on the stress tensor 
\ali{
	T_{X^+ X^+}^m(s) = T_{X^- X^-}^m(s).  
} 
Second, any ambiguity in the choice of integration limit ($u^\pm$ in the notation of \cite{Almheiri:2014cka}) can be absorbed in a redefinition 
of the integration constants in $\Phi_0$ (we are thus free to choose $u^+=u^-$ in the notation of \cite{Almheiri:2014cka}).

\section{Iyer-Wald formalism} \label{IWappendix}

We recall the Iyer-Wald (IW) formalism, following the notation of \cite{Iyer:1994ys}. 
Given a Lagrangian $L(\phi, \p \phi)$, one defines the energy variation through a region $\Xi$ as  
\ali{ 
	\delta H_\xi := \int_{\Xi} \omega(\phi, \delta \phi, \delta_\xi \phi) 
} 
in terms of the symplectic current 
\ali{
	\omega(\phi, \delta \phi, \delta_\xi \phi) &= \delta J_\xi + \xi \cdot E \, \delta \phi -  d(\xi \cdot \theta(\phi,\delta\phi)), 
}
where $E$ denotes the equations of motion for
the dynamical fields 
and $\theta$ the symplectic potential ($\delta L = E \delta \phi + d \theta$). 
$J_\xi$ is the current associated with the invariance of the Lagrangian under diffeomorphisms $\xi$, and is conserved on-shell. 
The corresponding Noether charge $Q_\xi$ is defined as  
\ali{
	J_\xi = d Q_\xi + C_\xi(E)   \label{defQ}
}
with $C_\xi(E)$ the constraint equations, such that $d J_\xi = -E \, \delta_\xi \phi$ \cite{Iyer:1996ky}. 
An IW first law is obtained when evaluating the relation 
\ali{ 
	\delta H_\xi :=  \int_\Xi \omega(\phi, \delta \phi, \delta_\xi \phi) &= \int_\Xi
	\delta J_\xi + \xi \cdot E \, \delta \phi -  d(\xi \cdot \theta(\phi,\delta\phi)) \label{applytoJT}
}
for a particular choice of diffeomorphism $\xi$, usually a Killing vector ($\delta_\xi \phi = 0$ for gravitational fields $\phi = \phi_g$, so that the symplectic current, which is bilinear in the variations, vanishes). 

Now let us apply the IW formalism to a gravitational theory with an action that depends on dynamic gravitational fields $\phi_g$ (including the metric) and matter fields $\phi_m$, for a region $\Xi$ and a vector 
$\xi$ that obeys $\xi |_{\p \Xi} = 0$. 
On a solution, $E=0$, and as $\xi |_{\p \Xi} = 0$,  
\ali{
	\delta H_{\xi}  &= \int_\Xi \delta J_{\xi}.  
}
Following the partition of the action in a gravitational and a matter part, the left hand side splits in a gravitational part $\omega_g(\phi_g, \delta \phi_g, \delta_\xi \phi_g)$ and a matter part $\omega_m(\phi_g,\phi_m,\delta \phi_g, \delta \phi_m, \delta_\xi \phi_g,$ $\delta_\xi \phi_m)$. The right hand side can be rewritten making use of \eqref{defQ} and assuming the variation $\delta \phi$ is to a nearby solution (so that $\delta d Q = d \delta Q$), obtaining  
\ali{
	\delta H_{\xi}^g + \delta H_{\xi}^m &=  \int_{\p \Xi} \delta Q_{\xi} + \int_\Xi \delta C_{\xi}(E).  \label{IWfirstlaw}
} 
This expresses that the on-shell vanishing of the linearized constraint equations is equivalent to the on-shell identity 
\ali{
	\delta H_{\xi}^g + \delta H_{\xi}^m &=  \int_{\p \Xi} \delta Q_{\xi}.  \label{IWfirstlawonshell}
}

\bibliographystyle{JHEP}
\bibliography{referencesDraft1}

\end{document}